\documentclass[11pt]{article}
\usepackage{UF_FRED_paper_style}
\usepackage{graphicx}
\DeclareGraphicsExtensions{.pdf,.png,.jpg,.jpg}
\usepackage{lipsum}
\title{Becoming Green: Decomposing the Macroeconomic Effects of Green Technology News Shocks}

\author{Oscar Jaulín\\Carlos III University of Madrid \\
    \href{osjaulin@eco.uc3m.es }{\texttt{osjaulin@eco.uc3m.es }}
\and Andrey Ramos\\Carlos III University of Madrid \\
    \href{anramosr@eco.uc3m.es}{\texttt{anramosr@eco.uc3m.es}}
    }   
\date{\today}

\begin{document}
{\setstretch{1}
\maketitle
\begin{abstract}
\vspace{0.3 cm}
\noindent
This paper studies the macroeconomic effects of news about future technological advancements in the green sector. Utilizing the economic value of green patents granted to publicly listed companies in the U.S., we identify green technology news shocks via a convenient and meaningful rotation of the innovations from a Bayesian Vector Autorregresion Model (BVAR). These shocks are decomposed into two orthogonal components: \textit{i)} a common technological component shared by both green and non-green innovation, that reproduces response patterns similar to those expected from a technology news shock with long-run impacts on productivity; and \textit{ii)} an idiosyncratic component to green innovation inducing inflationary pressures and stock price reductions. The responses to the idiosyncratic component suggest the existence of a green transition news mechanism related to expectations of more rigorous carbon policies or stricter environmental standards in the future. The focus on green innovation deepens our understanding about the effect of technology-specific news shocks and provides information of practical importance for macroeconomic and environmental policies. 
\end{abstract}
}
\vfill
\newpage

\section{Introduction}

Green innovation is key in the fight against climate change. By making new low-carbon technologies (LCTs) available, green innovation is a powerful tool in curbing emissions and helping firms and households to adapt to the adverse impact of climate change. The macroeconomic implications of the development of new technologies are an under-explored topic in the empirical literature. From a path-dependency argument \citep{Acemogluetal2016, Aghionetal2016}, green innovation disrupts existing carbon-intensive economic systems by rendering current production processes obsolete. It can reduce potential productivity benefits in the short and medium run. Conversely, \cite{AmbecLanoie} argues that green innovation can boost investment and gradually enhance productivity by improving energy efficiency and reducing energy costs. Green innovation can also generate broader knowledge spillovers than its carbon-intensive counterparts, thus fostering overall innovation \citep{Dechezlepretre, Fried}. Yet the empirical evidence on what of these effects dominates is inconclusive.

In this paper, we empirically explore the macroeconomic effects of anticipations about future technological advancements in the green sector. Following the work of \cite{CascaldiGarcia} and \cite{MirandaAgrippino}, we explode the informational content about future technology contained in patents data. Utilizing the economic value of green patents granted to publicly listed companies in the U.S. as in \cite{KPSS}, we identify green technology news shocks via a convenient and meaningful orthogonalization of the reduced-form innovations from a Bayesian Vector Autorregresion Model (BVAR). These shocks are decomposed into two orthogonal components: \textit{i)} a common technological component shared by both green and non-green innovation; and \textit{ii)} an idiosyncratic component to green innovation. 

The responses of key macroeconomic variables to the two green news components reveal interesting insights. A shock to the common component of green-technology news triggers macroeconomic responses that closely resemble those produced by a standard technology-news shock with long-run productivity impacts, as described by \cite{CascaldiGarcia}. The impact response of utilization-adjusted Total Factor Productivity (TFP) is zero, while a significant positive effect is first obtained seven to eight quarters after the shock, and remains persistent throughout the examined horizon. In line with \cite{BarskySims}, the delay between the occurrence of the shock and its actual positive effect on productivity is indicative of the identified component as containing relevant information on future, rather than present, aggregate productivity levels. A positive impact response in output, consumption, investment, and worked hours is obtained with an effect that is short-lived. By contrast, a shock to the idiosyncratic component produces a pronounced and persistent increase in the consumer price index along with a sustained decline in stock prices. These responses align with viewing the idiosyncratic component as containing forward-looking information about the green transition, specifically in the form of tighter climate policy anticipated for the future.

This paper is related to two strands of literature. First, the empirical literature on the identification of technology news shocks and their macroeconomic impacts. While existing literature predominantly focuses on aggregate technology news, our focus on green technology-specific news offers a deeper understanding on the impacts depending on the nature of the shocks and provides relevant information for policy decisions. Second, the literature on the economic benefits and costs associated with the green transition. Most of the evidence stems from simulation results derived from large-scale, general equilibrium models. Our contribution enhances the discussion by providing empirical estimates that complement previous works as \cite{Metcalf}, \cite{MetcalfStock}, or \cite{Kanzing}. A related study by \cite{Hasnaetal} at the IMF also examines the macroeconomic impacts of green innovation. The key difference is that, unlike their approach, we do not treat patent filings as indicators of actual innovation but rather as signals of future technological developments, in line with \cite{BeaudryPortier}. Additionally, our decomposition of shocks into components reflecting technological developments and the green transition represents an innovative contribution to this literature.

The rest of the paper is organized as follows. Section \ref{related_literature} describes the related literature and the contribution of our paper. Section \ref{data} presents the patent data sources and the construction of patent-based innovation indexes, as well as the macroeconomic data sources. Section \ref{methodology} introduces the methodology and the identification and decomposition of shocks in the context of a BVAR model. Empirical results are described in Section \ref{empirical}. Finally, Section \ref{conclusion} concludes.

\section{Related Literature} \label{related_literature}

Our paper relates to the empirical literature on identifying technology news shocks and their macroeconomic impacts. Since the seminal work of \cite{BeaudryPortier}, a substantial body of research has explored the role of advanced information about technological improvements in driving business cycle fluctuations. A key aspect of this literature is the identification assumptions. Traditional strategies, grounded in economic theory, often combine zero restrictions on the immediate response of TFP with assumptions about the long-run drivers of productivity. For example, \cite{BeaudryPortier} identify technology news shocks as innovations to the stock market index that are orthogonal to current TFP levels. Similarly, \cite{BarskySims} assume that technology news shocks are orthogonal to current TFP but maximize the forecast error variance at all horizons from zero to forty quarters. \cite{Francisetal} propose a related approach, imposing a zero restriction on the immediate response of TFP while assuming that news shocks maximize TFP's forecast error variance at a forty-quarter horizon. \cite{KurmannSims} adopt the methodology of \cite{Francisetal} but without the zero-impact restriction. In all cases, the TFP response to news shocks is imposed by the identification strategy rather than emerging as an independent empirical result.

In a pair of recent studies, \cite{CascaldiGarcia} and \cite{MirandaAgrippino} implement alternative identification strategies that do not impose particular dynamics on the TFP response. Both contributions are based on the premise that patents constitute a relevant source of information on inventive activity and signal potential future technological advancements. \cite{MirandaAgrippino} constructs a proxy for technological news shocks built from the number of quarterly patent applications filed at the USPTO. \cite{CascaldiGarcia} adds to the literature by exploiting the stock market valuations of patents granted to publicly listed firms, as captured by the innovation index proposed by \cite{KPSS}. Despite not being explicitly assumed, both papers obtain a shape for the TFP response that aligns with the conceptualization of \cite{BeaudryPortier}. Our study extends the methodology of \cite{CascaldiGarcia} by exploring the identification of technology-specific news shocks. While existing literature predominantly focuses on aggregate technology news, our focus on green technology-specific news offers a deeper understanding on the impacts depending on the nature of the shocks and provides relevant information for policy decisions. 

This paper is also linked to the literature on the economic benefits and costs associated with the green transition. In the long-run, the positive benefits of a cleaner economy in terms of mitigating climate change damages are widely recognized, see \cite{Acemogluetal2012}. Nonetheless, the transition towards a green economy, prompted by carbon taxes and/or subsidies to green investments, involves a complex array of short- to medium-run economic implications of practical importance for macroeconomic and environmental policies. For instance, \cite{Goulder} models the economic impact of carbon taxation and estimates that that implementing a carbon tax of 40 dollars per ton starting in 2020 and rising at 5 percent real annually,  would result in an output reduction of just over 1 percent by 2035 compared to a scenario without such a tax.

Other group of authors have used New Keynesian frameworks to explore the monetary impacts of transition policies. \cite{Ferrari2022}, for instance, finds that a current increase in a carbon tax exerts inflationary pressures, whereas anticipated future increases dampen demand. Expectations play a crucial role in determining the prevailing effect. \cite{Airaudoetal}, in a model for a small open economy, observe that increases in brown energy taxation lead to a rise in firms' marginal costs, with inflationary effects and persistent output losses. Green public investment or subsidies would induce a transition with no inflationary or output costs, albeit without a quick improvement in energy efficiency. \cite{DelNegro} point out that climate policies introduce an inflation-output trade-off, modulated by the relative price  stickiness in the dirty and green sectors and whether the policies involve taxes or subsidies. Other relevant references along this line include \cite{Bartocci} and \cite{Ferrari2023}.

Most of the evidence on the real and monetary consequences of the green transition stems from simulation results derived from large-scale, general equilibrium models. From an empirical perspective, recent contributions examine the macroeconomic effects of carbon taxation, albeit with conflicting findings. \cite{Metcalf} and \cite{BernardKichian}, despite adopting different methodological frameworks, conclude that the British Columbia carbon tax does not detrimentally affect output or employment. This view is further supported by \cite{MetcalfStock} who find no significant adverse effects on employment or GDP growth of carbon taxation in various European countries. In contrast, \cite{Kanzing} documents that a carbon policy shock in Europe leads to higher energy prices, lower emissions, and more green innovation, at the cost of a fall in economic activity.

Our analysis contributes to the empirical literature by examining a complementary and under-explored aspect of the green transition: the macroeconomic implications of news about future technological advancements in the green sector. Various mechanisms through which expected and actual green innovation influences economic decisions have been proposed. From a path-dependency argument \citep{Acemogluetal2016, Aghionetal2016}, green innovation disrupts existing carbon-intensive economic systems by rendering current production processes obsolete. In the short and medium run, potential productivity benefits are reduced. Conversely, \cite{AmbecLanoie}suggest that green innovation can boost investment and gradually enhance productivity by improving energy efficiency and reducing energy costs. In the same direction,  green innovation can generate broader knowledge spillovers than its carbon-intensive counterparts, thus fostering overall innovation \citep{Dechezlepretre, Fried}. Our empirical investigation, using U.S. data and a transparent strategy for identifying and decomposing green technology news shocks, addresses this topic. 

A related work is the study by \cite{Hasnaetal} at the IMF, which examines the macroeconomic and firm-level impacts of green innovation in the short and medium run. Drawing on data on patent fillings in OECD and BRICS countries from 1990 to 2020, these authors find that green patent filings stimulate output through increased investment, yet do not boost aggregate TFP. At the firm level, green patents increase revenue, albeit in a smaller magnitude compared to non-green patents. 

Our study deviates from  \cite{Hasnaetal} in several aspects. First, we do not conceive patent fillings as an indicator of actual innovation but rather as news about future technological developments in the sense of \cite{BeaudryPortier}. Second, we employ a transparent identification strategy that does not rely on the use of instrumental variables for addressing potential endogeneity issues.\footnote{Patent applications may be prompted by current economic booms and/or past news.} Our strategy uses the market values of the granted patents—instead of patent fillings—to construct the green and non-green innovation indexes in the spirit of \cite{KPSS}. It provides our innovation indicator with an additional degree of exogeneity coming from the high-frequency variation in asset prices. Technology news shocks are identified within a VAR model assuming convenient meaningful rotations of the reduced-form innovations. Third, we decompose the shocks into information regarding technological developments and information about the green transition. To the best of our knowledge, this is the first paper proposing this type of decomposition.

\section{Data} \label{data}

\subsection{Patent Data and Green Patent-Based-Innovation-Index (GPBII)} 

Data on patents is obtained from \textit{Patents View}, a publicly accessible service maintained by the U.S. Patent and Trademark Office (USPTO). For each patent granted between 1960 and 2016, we collect information on the application and grant dates and the technology class indicated by the Cooperative Patent Classification (CPC) code. Based on the CPC code and following the classification system proposed by \cite{HascicMigotto}, we identify the green patents as all patents producing technologies related to climate change mitigation and adaptation, carbon capture and storage, renewable energy generation, pollution abatement, and waste management.

\begin{figure}[H]
\begin{subfigure}{.33\textwidth}
  \centering
  \includegraphics[width=.95\linewidth]{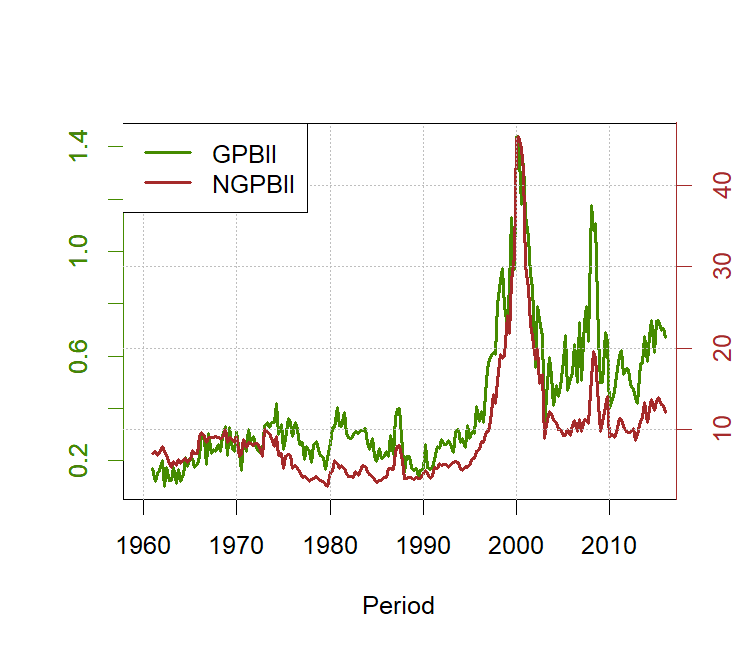}
  \caption{Level}
\end{subfigure}
\begin{subfigure}{.33\textwidth}
  \centering
  \includegraphics[width=.95\linewidth]{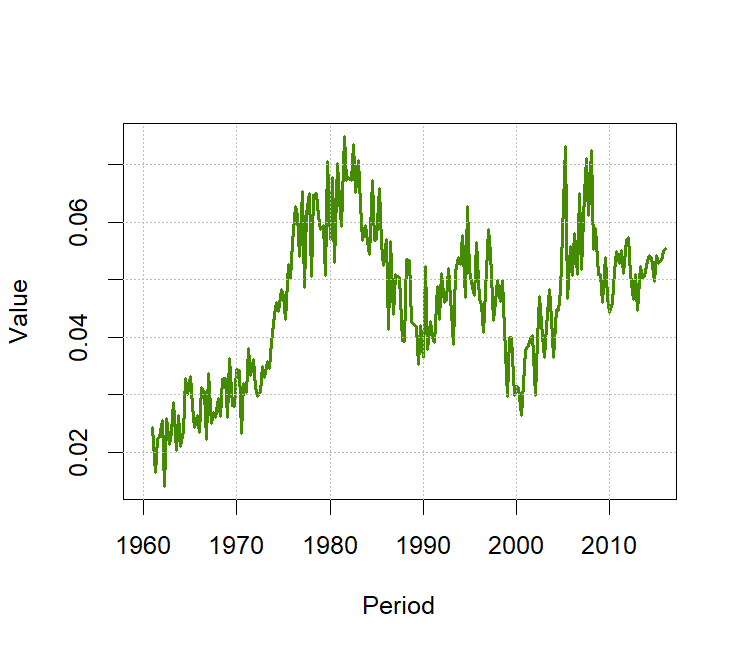}
  \caption{Ratio}
\end{subfigure}
\begin{subfigure}{.33\textwidth}
  \centering
  \includegraphics[width=.95\linewidth]{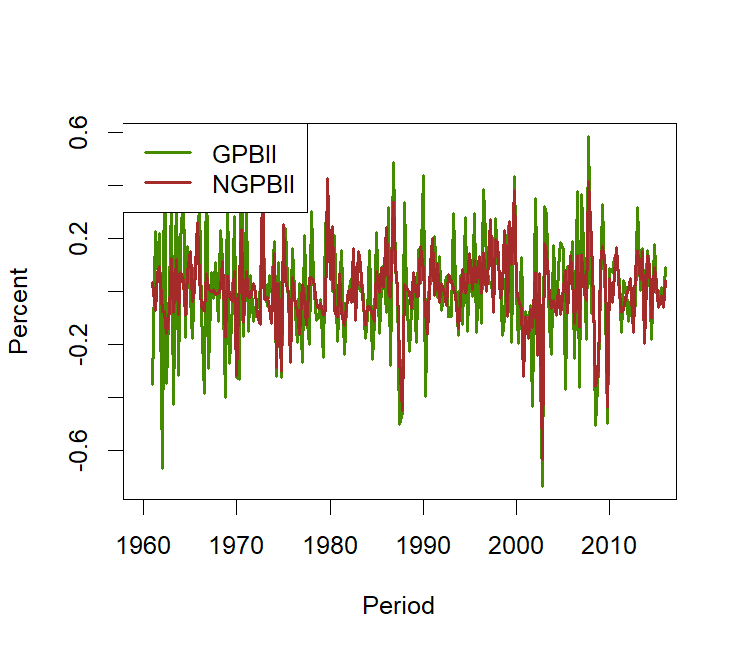}
  \caption{Growth rate}
\end{subfigure}%
\caption{Green and non-green PBIIs.}
\label{fig_index_0}
\end{figure}

Quarterly data on patents grants is combined with firm-level data from the Centre for Research in Security Prices (CRSP) to construct the Green Patent-Based-Innovation-Index (GPBII) index, following the methodology of \cite{KPSS}.\footnote{Codes and data files available at the GitHub repository \url{https://github.com/KPSS2017} were used.} Shortly, the value of an individual green and non-green patent granted to a publicly listed firm is estimated by filtering the firm's stock price reaction around the patent grant announcement date from other unrelated news. Individual firm-level values are added up each quarter to obtain the indexes presented in Figure \ref{fig_index_0}. 

Panel (a) in Figure \ref{fig_index_0} presents the levels of the GPBII (green) and NGPBII (brown). Both indices seem to evolve following a common trend. From 1990, both indices report a steeper increase that is interrupted at the beginning of the 2000s, with a more marked fall in the NGPBII. The value of the indices seems to follow times of speculation in the market, especially that of the dot-com bubble. In the mid-2000s, the GPBII appears to increase more steadily, but again a strong fall is observed around 2008, during the Great Recession. Even though the variability of both indices is similar—the correlation coefficient of the levels is above 0.83—the levels themselves are not. The ratio presented in Panel (b) shows that the GPBII is around 5\% of the NGPBII. However, during the sample period, this ratio has varied significantly, experiencing periods of increase between 1960 and 1982 or 2000 to 2008, where the maximum peaks of the ratio were achieved. Panel (c) plots together the growth rates. The correlation of this measure is around 0.6.

\subsection{Aggregate Macroeconomic Data}

For the aggregate analysis, we use data on a range of variables related to technology, real macroeconomics, and forward-looking indicators. The information set in our benchmark model includes the GPBII and NGPBII, utilization-adjusted TFP \citep{Fernald}, output, real consumption, real investment, hours worked, consumer price index, federal funds rates, consumer confidence, and stock prices. These variables are chosen as to encompass the set of variables used in the analysis of CG-V and \cite{MirandaAgrippino}, and to cover a wide spectrum of the macroeconomy. All variables are in log levels, following \citep{sims1990}, to account for possible cointegration. The data frequency is quarterly and covers the period from 1961:Q1 to 2016:Q4, which is the time window with more reliable data on patents. In the Appendix, we provide more details about the data sources and transformations.

\section{Methodology and Identification} \label{methodology}

We estimate a BVAR model with standard Normal-Inverse-Wishart (NIW) distribution priors. The choice of a BVAR is motivated by its ability to address the curse of dimensionality found in large VAR models, which is particularly relevant in our study. This is because we define a VAR system with more than 10 endogenous variables, an intercept, and four lags, and our data is sourced at quarterly frequency. The shocks of interest are obtained following the strategy in CGV. Specifically, we use a short-run identification strategy by using the Cholesky decomposition of the variance-covariance matrix of the reduced form residuals, with the GPBII or the NGPBII ordered on top of the information set. The results are presented as median impulse responses to the identified structural shocks, with one-standard-deviation coverage bands provided for inference. 

After the initial analysis, we set our benchmark exercise by adopting a procedure that simultaneously utilizes both the GPBII and the NGPBII. This procedure entails decomposing the reduced-form residuals of the GPBII into a common component ($\epsilon_{C,t}^{G}$) and idiosyncratic component ($\epsilon_{O,t}^{G}$), both of them with meaningful interpretation. The impulse responses of the macroeconomic variables to each of these components are obtained using cumulative-type local projections \citep{Jorda} of the form:

\begin{gather}
    y_{t+h} - y_{t-1} = \alpha_{h} + \beta_{h} Shock_t + \varepsilon_{t+h},
\end{gather}

where $y_{t+h} - y_{t-1}$ represents the long difference of the macroeconomic variable of interest, $Shock_{t}$ is either $\epsilon_{C,t}^{G}$ or $\epsilon_{O,t}^{G}$, and $\beta_{h}$ is the impulse-response to be estimated. In the subsequent sections, we elaborate more on these procedures.

\section{Empirical Results} \label{empirical}

\subsection{Aggregate green technology news shocks}

In an initial empirical exercise, we identify the response of macroeconomic variables to green technology news shocks using the approach of CGV. Specifically, we place the GPBII at the top of the information set and achieve identification through a Cholesky decomposition of the reduced-form residuals in the estimated BVAR model.

\begin{figure}[H]
\begin{subfigure}{.33\textwidth}
  \centering
  \includegraphics[width=.95\linewidth]{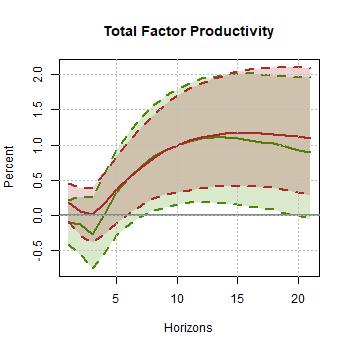}
\end{subfigure}
\begin{subfigure}{.33\textwidth}
  \centering
  \includegraphics[width=.95\linewidth]{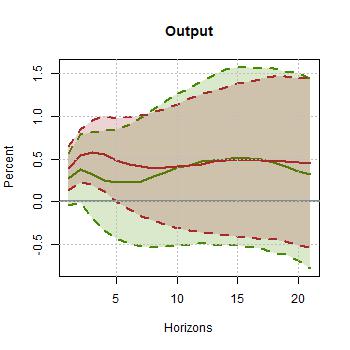}
\end{subfigure}
\begin{subfigure}{.33\textwidth}
  \centering
  \includegraphics[width=.95\linewidth]{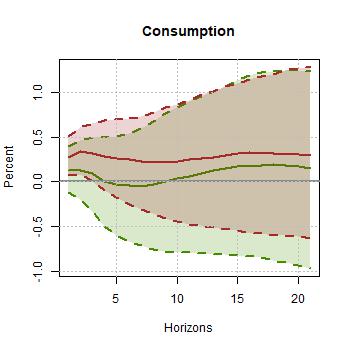}
\end{subfigure}%
\hfill
\begin{subfigure}{.33\textwidth}
  \centering
  \includegraphics[width=.95\linewidth]{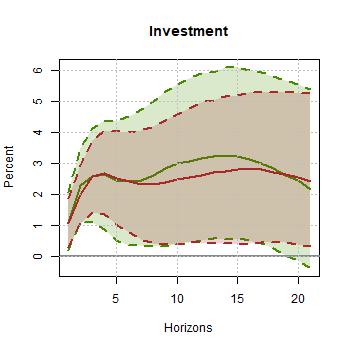}
\end{subfigure}
\begin{subfigure}{.33\textwidth}
  \centering
  \includegraphics[width=.95\linewidth]{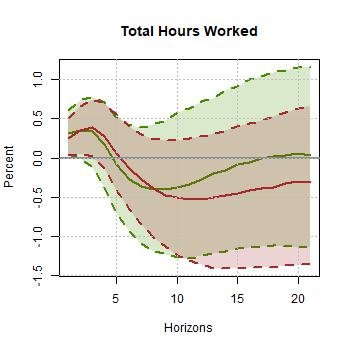}
\end{subfigure}
\begin{subfigure}{.33\textwidth}
  \centering
  \includegraphics[width=.95\linewidth]{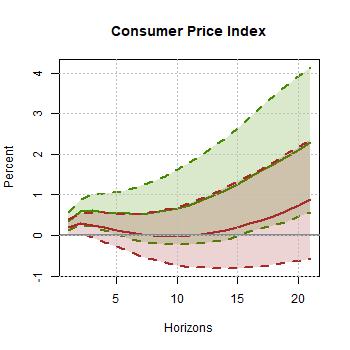}
\end{subfigure}%
\hfill
\begin{subfigure}{.33\textwidth}
  \centering
  \includegraphics[width=.95\linewidth]{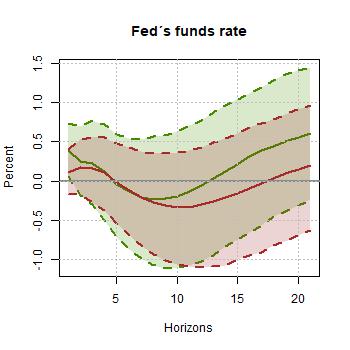}
\end{subfigure}
\begin{subfigure}{.33\textwidth}
  \centering
  \includegraphics[width=.95\linewidth]{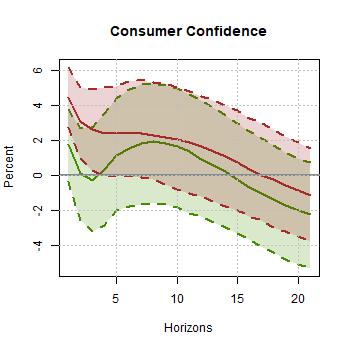}
\end{subfigure}
\begin{subfigure}{.33\textwidth}
  \centering
  \includegraphics[width=.95\linewidth]{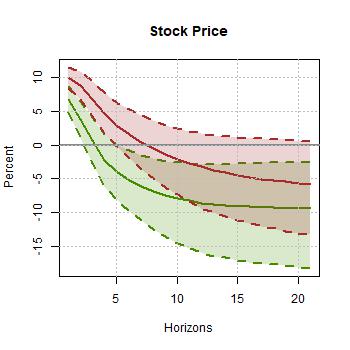}
\end{subfigure}
\caption{Dynamic responses to green (green) and non-green (brown) technology news shock identified in independent procedures.}
\label{fig_irf_0}
\end{figure}

This identification strategy assumes that the only structural shock affecting the contemporaneous value of the reduced-form residuals of the GPBII is the green technology news shock, which reflects expectations of future advancements in green technology. To ensure comparability, the shocks are rescaled so that the TFP increases by 1.0\% ten quarters after the shock. Green curves in Figure \ref{fig_irf_0} represent the implied dynamic responses of the macroeconomic variables to the identified shock over a twenty-quarter horizon, a duration deemed appropriate for future productivity forecasts. These responses are compared to the brown curves corresponding to the macroeconomic responses to a non-green technology news shock. Brown responses are obtained in a similar process but replacing the GPBII by the NGPBII as the first variable in the system. The latter analysis closely replicates the findings in CGV.\footnote{CGV use the total PBII, obtained as the sum of the GPBII and the NGPBII. Provided that the NGPBII significantly contributes to the total PBII, a strong similarity in the outcomes of the analysis is expected.}

The response patterns of green and non-green technologies exhibit notable similarities. In both cases, there is a significant lag between the initial impact of the technology news shock and its actual effect on TFP. This delay aligns with the expected dynamics of TFP adjustments to technology news shocks. Importantly, economic agents anticipate future productivity gains from technological advancements. This anticipation is reflected in the positive responses of output, investment, and hours worked, consistent with the findings of \cite{BeaudryPortier}.

However, is not the similarities but the important differences between each type that motivate our subsequent analyses. Notice that some macroeconomic variables have different responses between the two cases. The response patterns in the price index, federal funds rate, consumer confidence, consumption, and stock prices diverge from those of non-green news technology shocks.  In the green case, the price index response is positive and significantly different from zero twenty quarters after the shock. Regarding stock prices, although an initial increase is observed on impact, the response turns permanently negative from the fifth quarter onwards.

This analysis highlights important similarities in the responses of real economic and TFP variables to both identified shocks, suggesting that technological progress plays a role in both cases. This observation is further supported by the visual trends in the levels and growth rates of GPBII and NGPBII (Figure \ref{fig_index_0}), as well as the strong correlation of over 0.6 between the two shock series. 
While this commonality is relevant for policy analysis, our primary focus is on the idiosyncratic component of green technology news shocks, which may explain the distinct reactions of monetary and forward-looking variables. However, this idiosyncratic component cannot be independently identified. Instead, its isolation requires the simultaneous use of GPBII and NGPBII along with a transparent identification strategy, as outlined in the next section.

\subsection{Orthogonal decomposition of green technology news shocks} 

Green technology news shocks are identified through a joint procedure that integrates both NGPBII and GPBII into the information set. It is important to emphasize that green technology news shocks share a common component with non-green technology news shocks. To disentangle these components, we estimate a Bayesian Vector Autoregression (BVAR) model and extract the reduced-form residuals, $e_{t}^{NG}$ and $e_{t}^{G}$, associated with NGPBII and GPBII, respectively. Both residual series are correlated and, presumably, share a common component. Let $e^{NG}_t$ be the reduced-form residual of the NGPBII in the BVAR, $e^{G}_t$ is the residual of the GPBII and $\epsilon_{C,t}^{G}$ is the common component. We assume that the only driver of $e^{NG}_t$ is the common component, while there are two drivers of $e^{G}_t$. Formally:

\begin{gather*}
    e^{NG}_t = \epsilon_{C,t}\\
    e^{G}_t = \epsilon_{C,t}^{G} + \epsilon_{O,t}^{G},
\end{gather*}

implying that $e_{t}^{G}$ is decomposed into two orthogonal elements. The first element, denoted as $\epsilon_{C,t}^{G}$, is obtained as the projection of $e_{t}^{G}$ on the $e_{t}^{NG}$ space and represents the common component of technology news. The second element, $\epsilon_{O,t}^{G}$, is obtained as the residual of the previous projection and represents the idiosyncratic component of the green technology news. Both components are orthogonal and satisfy $e_{t}^{G} = \epsilon_{C,t} + \epsilon_{O,t}^{G}$. \footnote{In Appendix A.1 we consider one alternative to analyze the macroeconomic responses to the idiosyncratic component of the green technology news. The approach uses short-run restrictions of the BVAR with the second variable being the GPBII $e_{t}^{G}$ concerning $e_{t}^{NG}$. IRFs are obtained directly from the BVAR.}\footnote{It is easy to prove that impulse-responses to $e^{G}_t$ are a linear combination of the responses to $\epsilon_{C,t}^{G}$ and $\epsilon_{O,t}^{G}$.} The regression between the two reduced-form residuals yields a coefficient of determination ($R^2$) of 0.43, indicating that 43.0\% of the variance in green technology news shocks can be attributed to the common component shared with non-green technology news shocks.

\begin{figure}[htbp]
\begin{subfigure}{.33\textwidth}
  \centering
  \includegraphics[width=.95\linewidth]{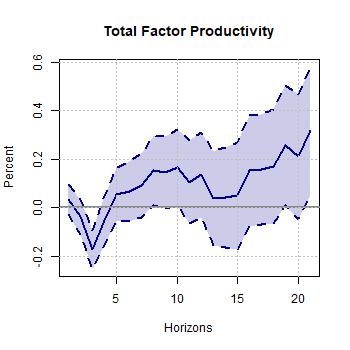}
\end{subfigure}
\begin{subfigure}{.33\textwidth}
  \centering
  \includegraphics[width=.95\linewidth]{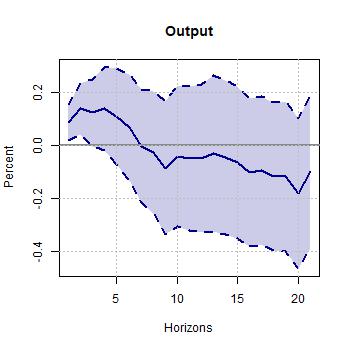}
\end{subfigure}
\begin{subfigure}{.33\textwidth}
  \centering
  \includegraphics[width=.95\linewidth]{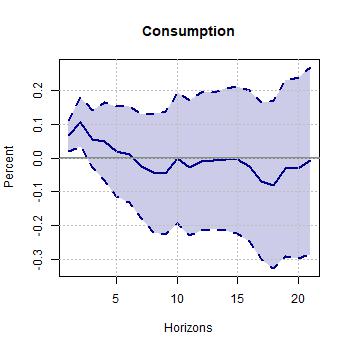}
\end{subfigure}%
\hfill
\begin{subfigure}{.33\textwidth}
  \centering
  \includegraphics[width=.95\linewidth]{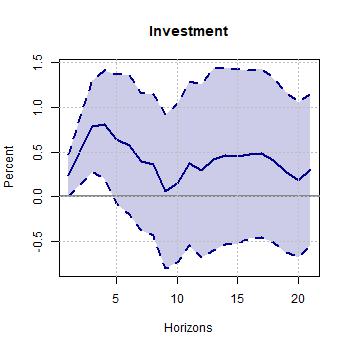}
\end{subfigure}
\begin{subfigure}{.33\textwidth}
  \centering
  \includegraphics[width=.95\linewidth]{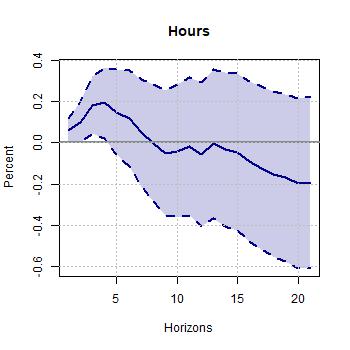}
\end{subfigure}
\begin{subfigure}{.33\textwidth}
  \centering
  \includegraphics[width=.95\linewidth]{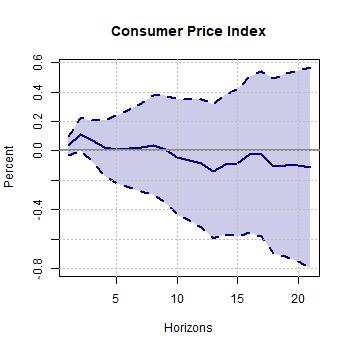}
\end{subfigure}%
\hfill
\begin{subfigure}{.33\textwidth}
  \centering
  \includegraphics[width=.95\linewidth]{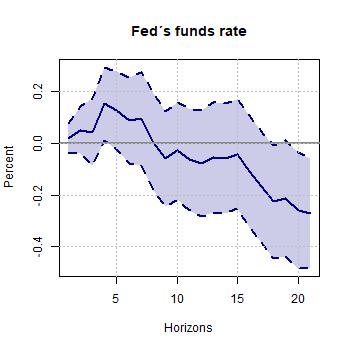}
\end{subfigure}
\begin{subfigure}{.33\textwidth}
  \centering
  \includegraphics[width=.95\linewidth]{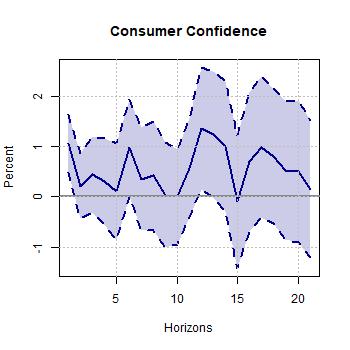}
\end{subfigure}
\begin{subfigure}{.33\textwidth}
  \centering
  \includegraphics[width=.95\linewidth]{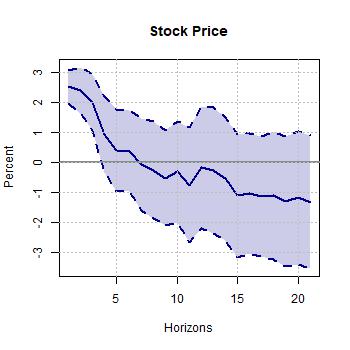}
\end{subfigure}
\hfill
\begin{subfigure}{.33\textwidth}
  \centering
  \includegraphics[width=.95\linewidth]{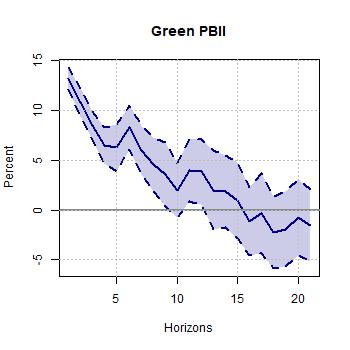}
\end{subfigure}
\begin{subfigure}{.33\textwidth}
  \centering
  \includegraphics[width=.95\linewidth]{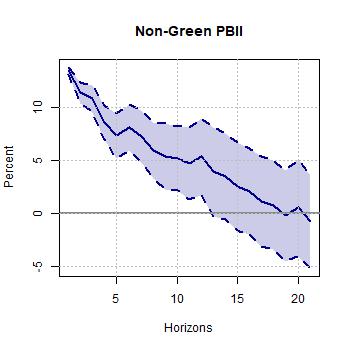}
\end{subfigure}
\caption{Dynamic responses to the common component of the green technology news shock (Baseline).}
\label{fig_irf_1}
\end{figure}

\begin{figure}[htbp]
\begin{subfigure}{.33\textwidth}
  \centering
  \includegraphics[width=.95\linewidth]{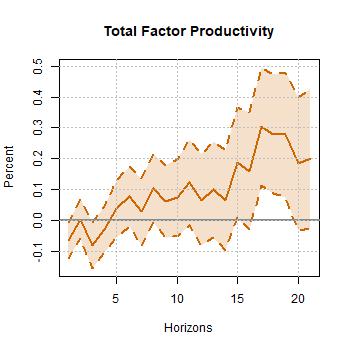}
\end{subfigure}
\begin{subfigure}{.33\textwidth}
  \centering
  \includegraphics[width=.95\linewidth]{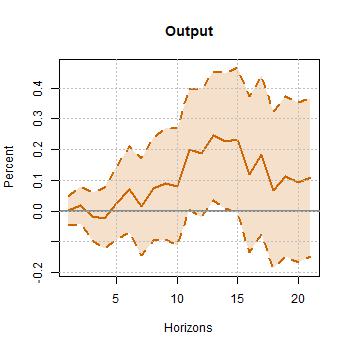}
\end{subfigure}
\begin{subfigure}{.33\textwidth}
  \centering
  \includegraphics[width=.95\linewidth]{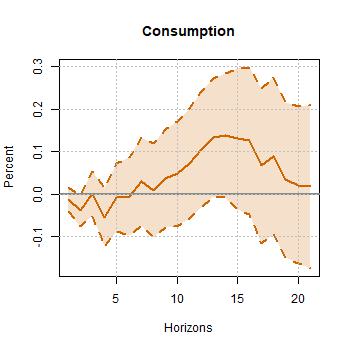}
\end{subfigure}%
\hfill
\begin{subfigure}{.33\textwidth}
  \centering
  \includegraphics[width=.95\linewidth]{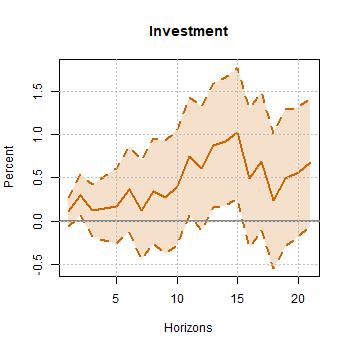}
\end{subfigure}
\begin{subfigure}{.33\textwidth}
  \centering
  \includegraphics[width=.95\linewidth]{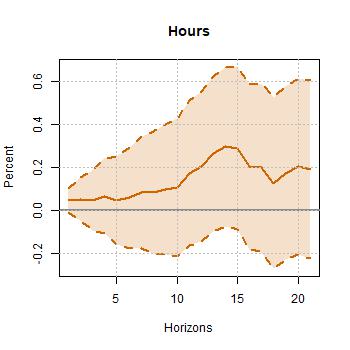}
\end{subfigure}
\begin{subfigure}{.33\textwidth}
  \centering
  \includegraphics[width=.95\linewidth]{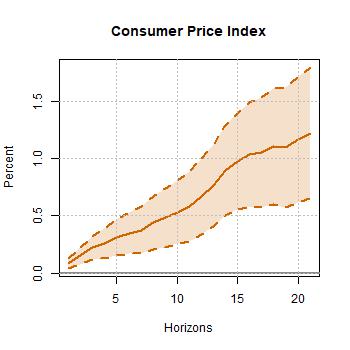}
\end{subfigure}%
\hfill
\begin{subfigure}{.33\textwidth}
  \centering
  \includegraphics[width=.95\linewidth]{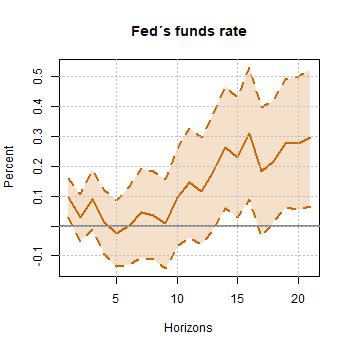}
\end{subfigure}
\begin{subfigure}{.33\textwidth}
  \centering
  \includegraphics[width=.95\linewidth]{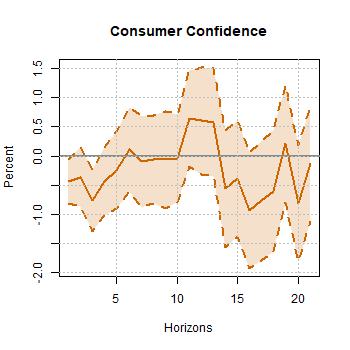}
\end{subfigure}
\begin{subfigure}{.33\textwidth}
  \centering
  \includegraphics[width=.95\linewidth]{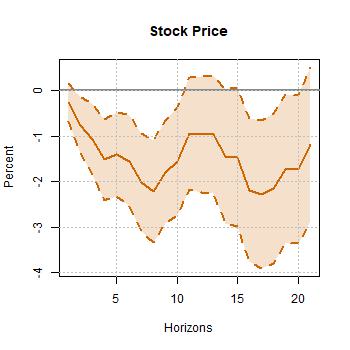}
\end{subfigure}
\hfill
\begin{subfigure}{.33\textwidth}
  \centering
  \includegraphics[width=.95\linewidth]{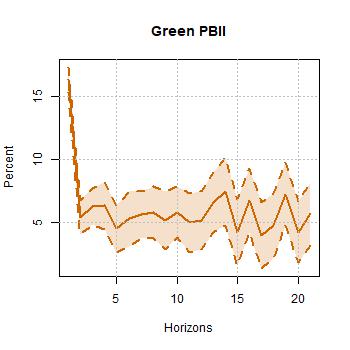}
\end{subfigure}
\begin{subfigure}{.33\textwidth}
  \centering
  \includegraphics[width=.95\linewidth]{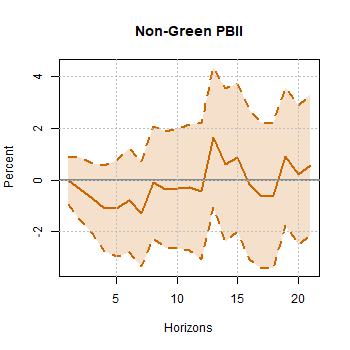}
\end{subfigure}
\caption{Dynamic responses to the idiosyncratic component of green technology news shocks (Baseline).}
\label{fig_irf_2}
\end{figure}

The dynamic responses of the macroeconomic variables to a one standard deviation increase in $\epsilon_{C,t}^{G}$ and $\epsilon_{O,t}^{G}$ are obtained using cummulative-type local projections \cite{Jorda}.\footnote{We estimate them by using OLS and HAC standar errors for the confidence intervals \citep{jorda2023}.} Figures \ref{fig_irf_1} and \ref{fig_irf_2} plot those responses over a horizon of 20 quarters. Both types of shock components induce a significant increase of the GPBII on impact, yet their subsequent response patterns are completely different. Based on the results, we interpret $\epsilon_{C,t}^{G}$ as representing a common technological content and $\epsilon_{O,t}^{G}$ as indicative of the green idiosyncratic component of the green technology news.

\paragraph{Responses to the common technological content of green technology news}

This section describes the responses of the macroeconomic variables to a unit increase in the common component news technology shock, $\epsilon_{C,t}^{G}$. Firstly, we examine the response of the utilization-adjusted TFP, a proxy for the technological advancement of the economy. The conventional identification of news technology shocks relies on zero and sign restrictions to isolate shocks that have no immediate impact on TFP while leading to future increases \citep{BeaudryPortier,BarskySims}. In contrast, we remain agnostic as in \cite{CascaldiGarcia} and \cite{MirandaAgrippino} and our approach does not impose restrictions on the immediate response of TFP nor in the sign response of its future values. Notably, the confidence intervals do not exclude the zero contemporaneous effect. A negative response is observed at horizon $h=3$ that may be related to Schumpeter's idea of creative destruction. A significant positive effect is first obtained seven to eight quarters after the shock and remains persistent throughout the examined horizon. In line with the conceptualization of news shocks in \cite{BarskySims}, the delay between the occurrence of the shock and its actual positive effect on productivity is indicative of the identified component as containing relevant information on future, rather than present, aggregate productivity levels.

Regarding real variables, we note a positive impact response in output, consumption, investment, and hours worked, with an effect that is short-lived. For instance, investment increases on impact and continues to rise for about 4 quarters before converging back to normal. Although the function is positive over the whole horizon, it is non-significant after quarter 6. Forward-looking indicators, as stock prices and consumer confidence, experience a positive strong initial increase that swiftly diminishes and turns non-significant by 4 quarters post-shock onwards. Notice that the positive responses of real variables and forward-looking indicators precede the significant impact on TFP, a finding that reveals the technological information component contained in the green technology shock that prompts economic agents to act upon the realization of the shock. Consumer prices exhibit a weak rise on impact that can be explained by the demand increase induced by the movement in the real variables.

From the described responses, it is possible to relate the common component $\epsilon_{C,t}^{G}$ to technological content in the form of news about future technological development. The macroeconomic implications are comparable to those described in CGV and the related literature on technological news shocks.

\paragraph{Responses to the idiosyncratic transition content of green technology news}

Now, we describe the impulse responses of the macroeconomic variables to a unitary increase in $\epsilon_{O,t}^{G}$ that are plotted in Figure \ref{fig_irf_2}. Observe that such movement generates a significant immediate rise in the GPBII, whereas the NGPBII remains unchanged at any horizon. This outcome is a consequence of the orthogonalization process and implies that the responses of other variables in the model are driven by factors not correlated with fluctuations in the non-green innovation. Unlike the response to the technological component discussed in the previous section, the dynamics of the GPBII in the present case show an interesting pattern: after the initial rise, the GPBII eventually stabilizes at a level approximately 5\% above the starting point. 

The most striking result in Figure \ref{fig_irf_2} is the significant and persistent increase in the consumer price index (CPI). While the immediate effect is modest, the impulse response function exhibits a monotonic upward trend over the entire horizon. This finding can be interpreted by considering that $\epsilon_{O,t}^{G}$ incorporates relevant information about the green transition. As discussed in Section 5.3, green innovation has the potential to shape future climate policy.

If technological advancements emerge today, regulators may adjust future taxes and environmental standards to facilitate the integration of these innovations into production processes. Consequently, when economic agents receive unexpected news about future green technology developments, they anticipate potential changes in climate policy, which in turn influence present economic outcomes. The sustained rise in inflation, for instance, can be attributed to expected increases in energy prices: as dirty energy becomes relatively more expensive, overall price levels rise.

Moreover, this perspective may also explain the persistent impact of $\epsilon_{O,t}^{G}$ on green innovation, considering the well-documented role of climate policy in driving technological advancements in the literature.

The monetary policy stance exhibits a mild increase on impact, and its impulse response function shows an increasing behavior over the entire horizon, consistent with the rise in consumer prices. However, the point responses are barely significant, reflecting the policy trade-off introduced by the inflationary pressures. The weakly positive response aligns with the notion that the Central Bank adopts a monetary policy reaction function, placing positive weight on inflation and taking into account the economic consequences of the green transition. Regarding financial markets, it is observed that the stock market experiences a significant downturn following an increase in $\epsilon_{O,t}^{G}$. Stock prices reduce on impact and remain lower throughout the entire horizon, although the impulse response becomes statistically insignificant after 10 quarters post-shock. This pattern again can be explained by changes in expectations regarding a more stringent future climate policy. The responses of the real variables are not significant, although a U-shaped pattern is observable for output, consumption, investment, and hours worked, with the peak response occurring approximately 13 to 14 quarters after the shock. For the TFP, the peak response is observed at the 17th quarter and reaches statistical significance.

\begin{figure}[htbp]
\begin{subfigure}{.33\textwidth}
  \centering
  \includegraphics[width=.95\linewidth]{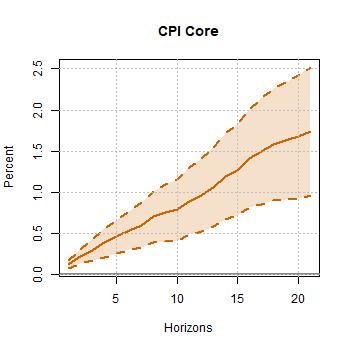}
\end{subfigure}
\begin{subfigure}{.33\textwidth}
  \centering
  \includegraphics[width=.95\linewidth]{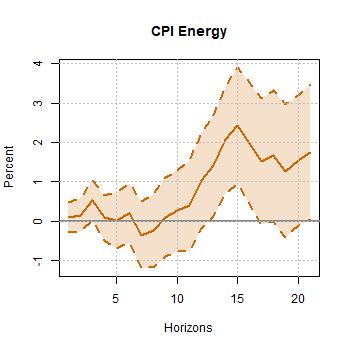}
\end{subfigure}
\begin{subfigure}{.33\textwidth}
  \centering
  \includegraphics[width=.95\linewidth]{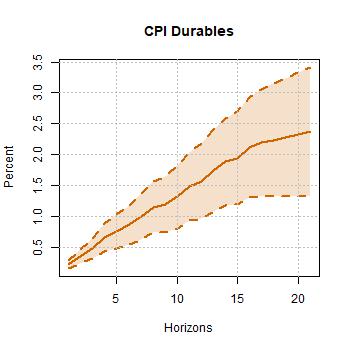}
\end{subfigure}%
\hfill
\begin{subfigure}{.33\textwidth}
  \centering
  \includegraphics[width=.95\linewidth]{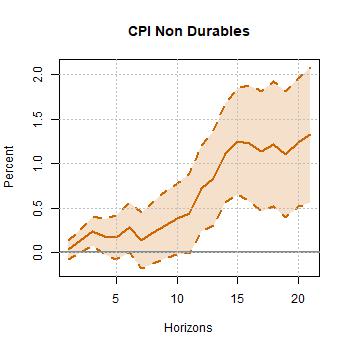}
\end{subfigure}
\begin{subfigure}{.33\textwidth}
  \centering
  \includegraphics[width=.95\linewidth]{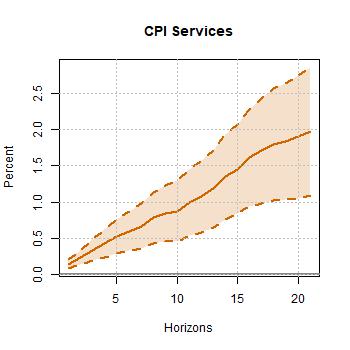}
\end{subfigure}
\caption{Dynamic responses to the idiosyncratic transition content of green technology news (CPI components).}
\label{fig_irf_2_prices}
\end{figure}

\textit{Consumer Prices.}—The idiosyncratic component of the green technology news shocks results in a strong and persistent increase in consumer prices. What are the responses of the energy prices and other specific price categories to these components? In Figure \ref{fig_irf_2_prices} we observe that the pattern response of core prices mirrors that of the overall price index in magnitude and shape. In contrast, energy consumer prices do not react in the short run, and a significant response is only observed around the 15th quarter following the shock. The responses obtained for the prices of durable goods and services closely resemble those of core CPI. Meanwhile, the price response of non-durable goods parallels that of energy prices.

\begin{figure}[htbp]
\begin{subfigure}{.33\textwidth}
  \centering
  \includegraphics[width=.95\linewidth]{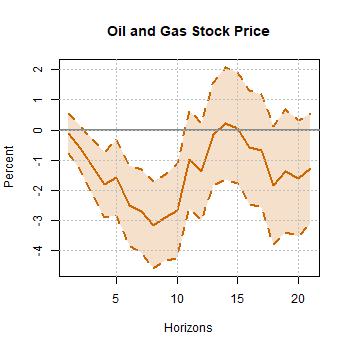}
\end{subfigure}
\begin{subfigure}{.33\textwidth}
  \centering
  \includegraphics[width=.95\linewidth]{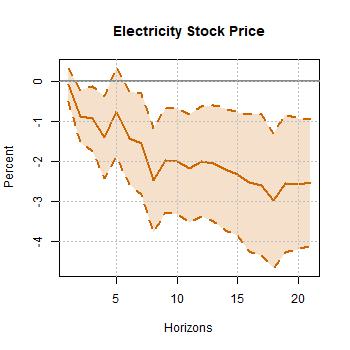}
\end{subfigure}
\begin{subfigure}{.33\textwidth}
  \centering
  \includegraphics[width=.95\linewidth]{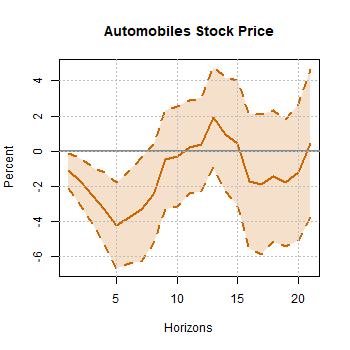}
\end{subfigure}%
\hfill
\begin{subfigure}{.33\textwidth}
  \centering
  \includegraphics[width=.95\linewidth]{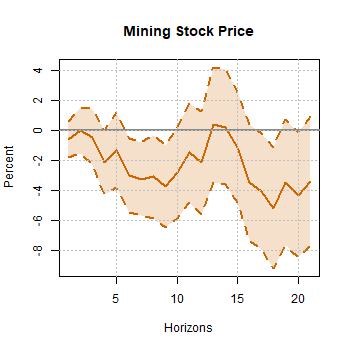}
\end{subfigure}
\begin{subfigure}{.33\textwidth}
  \centering
  \includegraphics[width=.95\linewidth]{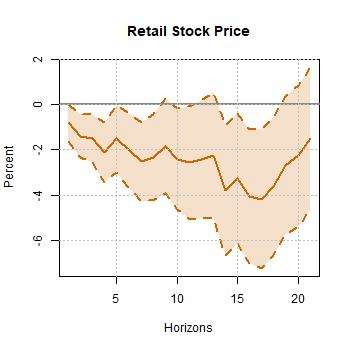}
\end{subfigure}
\begin{subfigure}{.33\textwidth}
  \centering
  \includegraphics[width=.95\linewidth]{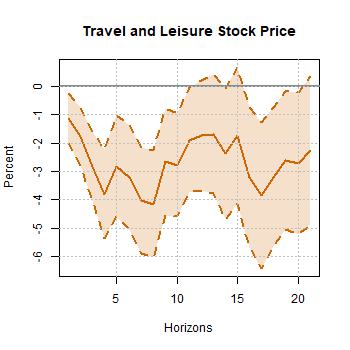}
\end{subfigure}
\caption{Dynamic responses to the idiosyncratic transition content of green technology news (Stock prices by industry).}
\label{fig_irf_2_stock}
\end{figure}

\textit{Stock Prices.}—To better understand the decline in aggregate stock prices, Figure \ref{fig_irf_2_stock} examines the stock price responses of selected industries potentially impacted by changes in environmental regulations. Except for the mining sector, all analyzed industries experienced a reduction in their stock prices. For the oil and gas and automobile sectors, the reduction is transitory, with the response becoming non-significant from the 10th horizon onwards, mirroring the dynamics of the aggregated stock price index. In contrast, the electricity and retail sectors exhibit a more persistent fall in their stock prices. Observed responses align with the green technology news as containing a transition content, which carries implications of future changes in environmental regulations and the necessary investments required to adapt to upcoming technological scenarios.

\subsection{The role of climate policy}

\begin{figure}[htbp]
\begin{subfigure}{.33\textwidth}
  \centering
  \includegraphics[width=.95\linewidth]{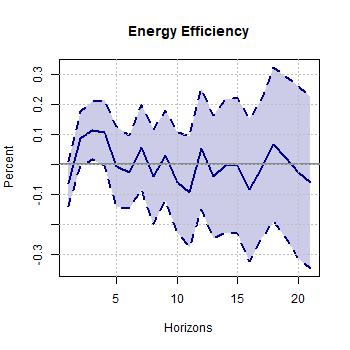}
\end{subfigure}
\begin{subfigure}{.33\textwidth}
  \centering
  \includegraphics[width=.95\linewidth]{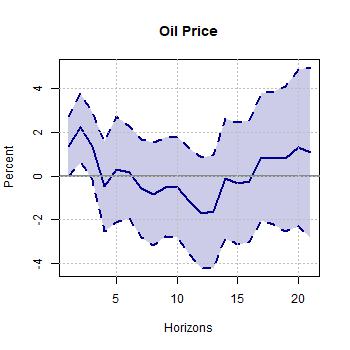}
\end{subfigure}
\begin{subfigure}{.33\textwidth}
  \centering
  \includegraphics[width=.95\linewidth]{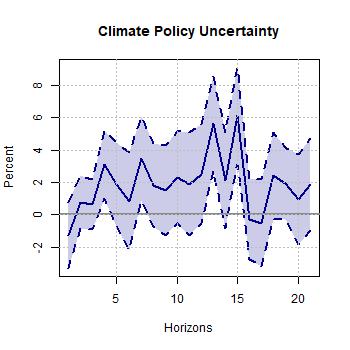}
\end{subfigure}
\begin{subfigure}{.33\textwidth}
  \centering
  \includegraphics[width=.95\linewidth]{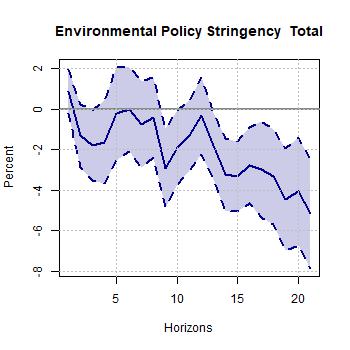}
\end{subfigure}
\begin{subfigure}{.33\textwidth}
  \centering
  \includegraphics[width=.95\linewidth]{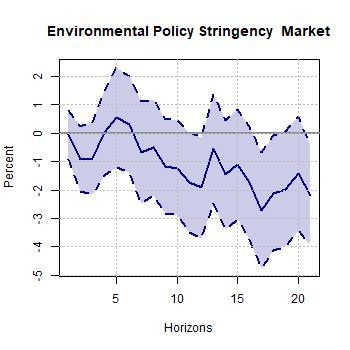}
\end{subfigure}
\begin{subfigure}{.33\textwidth}
  \centering
  \includegraphics[width=.95\linewidth]{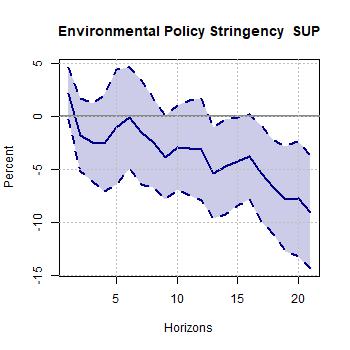}
\end{subfigure}%
\caption{Dynamic responses of climate policy variables to the common component of green technology news shocks.}
\label{fig_irf_4}
\end{figure}

\begin{figure}[htpb]
\begin{subfigure}{.33\textwidth}
  \centering
  \includegraphics[width=.95\linewidth]{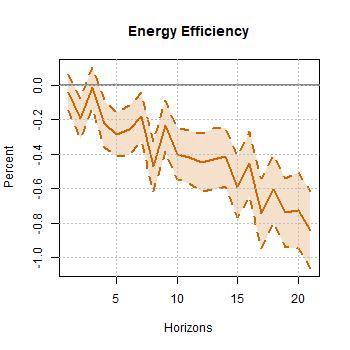}
\end{subfigure}
\begin{subfigure}{.33\textwidth}
  \centering
  \includegraphics[width=.95\linewidth]{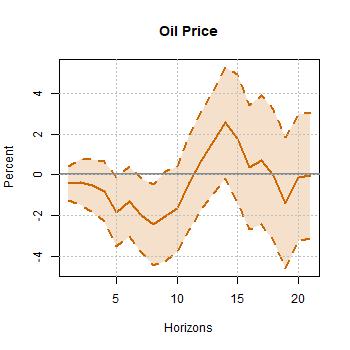}
\end{subfigure}
\begin{subfigure}{.33\textwidth}
  \centering
  \includegraphics[width=.95\linewidth]{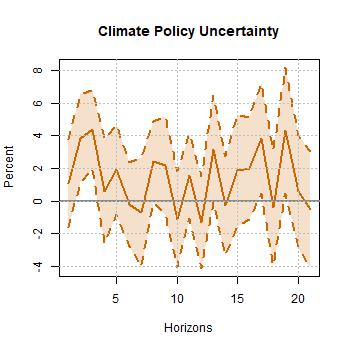}
\end{subfigure}
\begin{subfigure}{.33\textwidth}
  \centering
  \includegraphics[width=.95\linewidth]{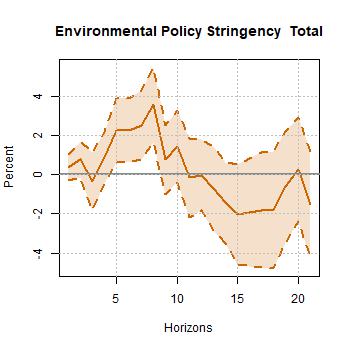}
\end{subfigure}
\begin{subfigure}{.33\textwidth}
  \centering
  \includegraphics[width=.95\linewidth]{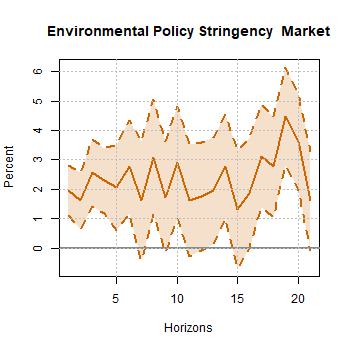}
\end{subfigure}
\begin{subfigure}{.33\textwidth}
  \centering
  \includegraphics[width=.95\linewidth]{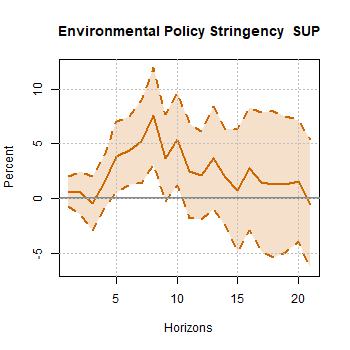}
\end{subfigure}%
\caption{Dynamic responses of climate policy variables to the idiosyncratic component of green technology news shocks.}
\label{fig_irf_5}
\end{figure}

The asymmetric responses of stock and consumer prices to the common and idiosyncratic components of green-technology news indicate that each component operates through a different set of transmission mechanisms. In this section, we argue that these mechanisms may be driven by two key factors: i) the energy efficiency, captured by fossil-fuel consumption as a share of GDP, and ii) the stringency and uncertainty of climate policy. Figures~\ref{fig_irf_4} and~\ref{fig_irf_5} display the local-projection impulse responses for these variables (methodology in Section~5.1.2).

Focus first in the responses of energy efficiency. The response of this indicator to the common component is not significant at any horizon, while the idiosyncratic component induces a permanent reduction. This permanent decline helps to explain the observed increase in the CPI: switching away from cheap fossil inputs toward costlier low-carbon alternatives raises the price level, a pattern consistent with the greenflation channel proposed by \citet{Airaudoetal}. Observe that this mechanism is not driven by higher energy prices since the WTI oil price shows no significant reaction to the idiosyncratic shock. Consistent with this observation, augmenting the BVAR with the energy-efficiency series neutralizes the CPI response, confirming the crucial role of this dimension of technology (see Figure~\ref{fig_irf_A2} in the Appendix).

Turning to climate policy, we track the responses of the Environmental Policy Stringency (EPS) index of \citet{botta} (reported for total, market-based, and technology support measures) and the Climate Policy Uncertainty (CPU) index of \citet{gavriilidis}. Following a shock to the idiosyncratic component of green technology, both CPU and all EPS variants rise significantly, with the strongest immediate jump in market-based instruments (carbon taxes, diesel taxes, CO2 trading schemes, etc.).  Investors plausibly interpret this shock as a signal of tougher future carbon regulation, depressing expected cash-flows and, hence, equity valuations. Firms may anticipate weaker future performance, prompting financial markets to respond with lower stock valuations and less favorable firm-level sales or profits. By contrast, shocks to the common component reduce the EPS indicator, especially its technology-support subindex, suggesting that broad, economy-wide technological optimism demands less additional policy push. 

\subsection{Time-varying effects after 1990}

A natural question is whether the two green-news orthogonal components propagate differently in the decades when environmental policy and clean-tech investment accelerated. To address this, we re-estimate the cumulative local-projection specification

\begin{gather}
    y_{t+h} - y_{t-1} = D_t\Big[\alpha_h(1) + \beta_h(1) shock_t\Big] + (1-D_t)\Big[ \alpha_h(0) + \beta_h(0) shock_t\Big] + \varepsilon_{t+h},
\end{gather}

where $D_{t} = I(t > 1990)$ is an indicator taking the value of 1 if the shock takes place after 1990, and $shock_t$ is either $\epsilon_{C,t}^{G}$ or $\epsilon_{O,t}^{G}$. The post-1990 dummy is motivated by the sharp rise in both PBII series that begins around 1990 and peaks in the early-2000s technological boom.

\begin{figure}[htp]
\begin{subfigure}{.33\textwidth}
  \centering
  \includegraphics[width=.95\linewidth]{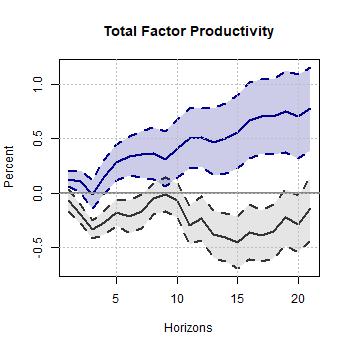}
\end{subfigure}
\begin{subfigure}{.33\textwidth}
  \centering
  \includegraphics[width=.95\linewidth]{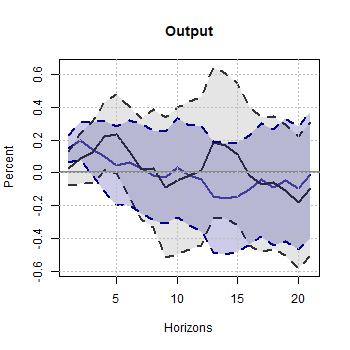}
\end{subfigure}
\begin{subfigure}{.33\textwidth}
  \centering
  \includegraphics[width=.95\linewidth]{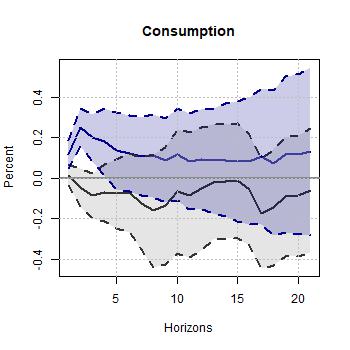}
\end{subfigure}%
\hfill
\begin{subfigure}{.33\textwidth}
  \centering
  \includegraphics[width=.95\linewidth]{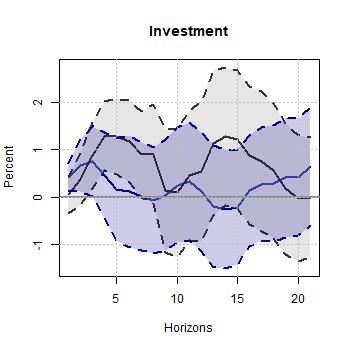}
\end{subfigure}
\begin{subfigure}{.33\textwidth}
  \centering
  \includegraphics[width=.95\linewidth]{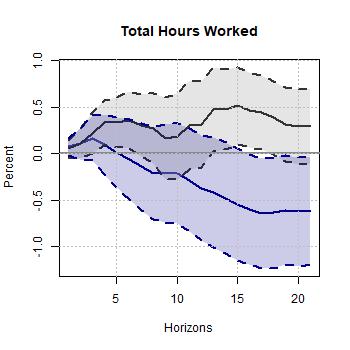}
\end{subfigure}
\begin{subfigure}{.33\textwidth}
  \centering
  \includegraphics[width=.95\linewidth]{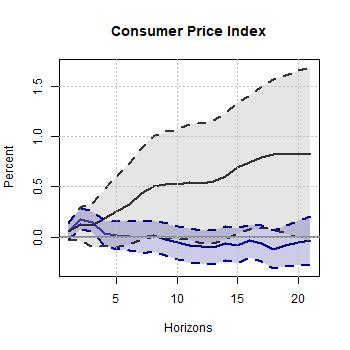}
\end{subfigure}%
\hfill
\begin{subfigure}{.33\textwidth}
  \centering
  \includegraphics[width=.95\linewidth]{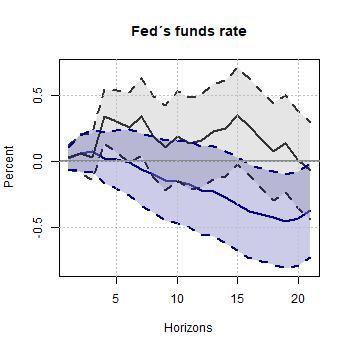}
\end{subfigure}
\begin{subfigure}{.33\textwidth}
  \centering
  \includegraphics[width=.95\linewidth]{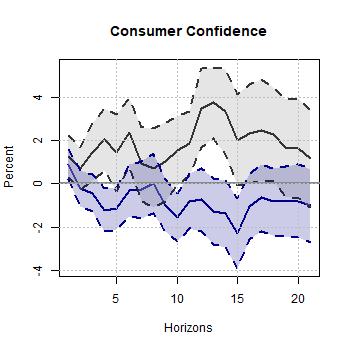}
\end{subfigure}
\begin{subfigure}{.33\textwidth}
  \centering
  \includegraphics[width=.95\linewidth]{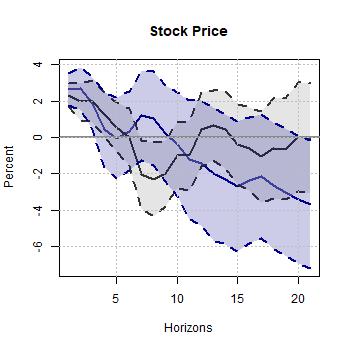}
\end{subfigure}
\hfill
\begin{subfigure}{.33\textwidth}
  \centering
  \includegraphics[width=.95\linewidth]{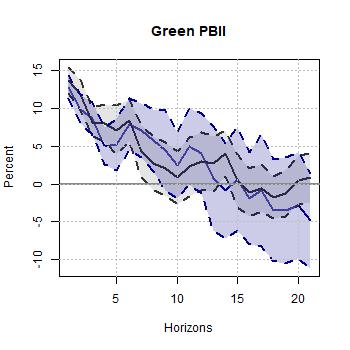}
\end{subfigure}
\begin{subfigure}{.33\textwidth}
  \centering
  \includegraphics[width=.95\linewidth]{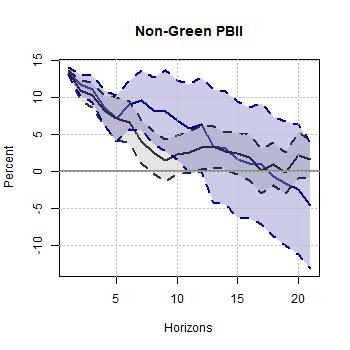}
\end{subfigure}
\caption{Dynamic responses to the common technological component of the green technology news shock. Responses before (gray) and after (blue) 1990 estimated respectively by $\beta_h(0)$ and $\beta_h(1)$ in Equation 2 using $\epsilon_{C,t}^{G}$ as the shock variable.}
\label{fig_irf_6a}
\end{figure}

\begin{figure}[htp]
\begin{subfigure}{.33\textwidth}
  \centering
  \includegraphics[width=.95\linewidth]{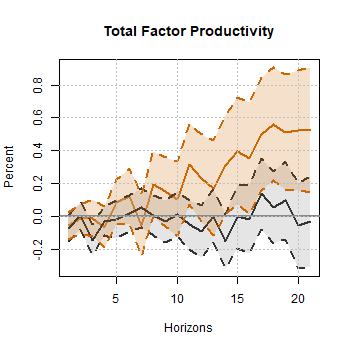}
\end{subfigure}
\begin{subfigure}{.33\textwidth}
  \centering
  \includegraphics[width=.95\linewidth]{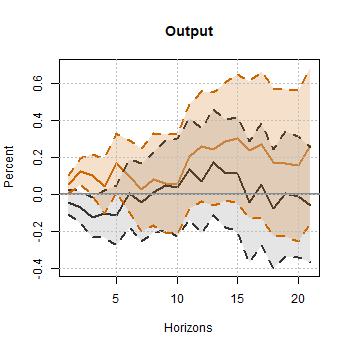}
\end{subfigure}
\begin{subfigure}{.33\textwidth}
  \centering
  \includegraphics[width=.95\linewidth]{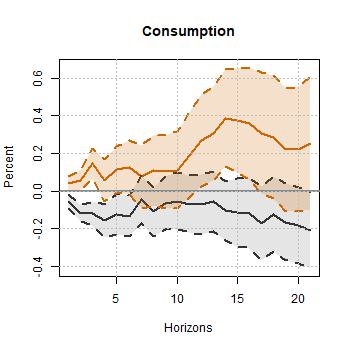}
\end{subfigure}%
\hfill
\begin{subfigure}{.33\textwidth}
  \centering
  \includegraphics[width=.95\linewidth]{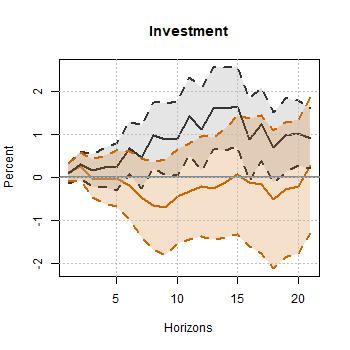}
\end{subfigure}
\begin{subfigure}{.33\textwidth}
  \centering
  \includegraphics[width=.95\linewidth]{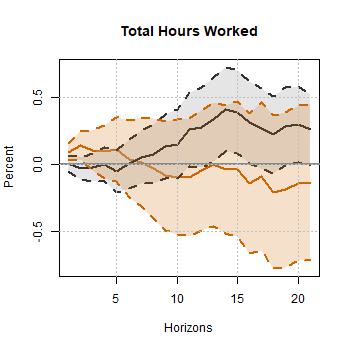}
\end{subfigure}
\begin{subfigure}{.33\textwidth}
  \centering
  \includegraphics[width=.95\linewidth]{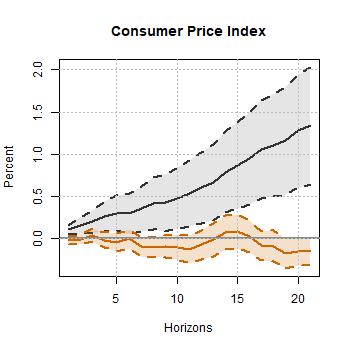}
\end{subfigure}%
\hfill
\begin{subfigure}{.33\textwidth}
  \centering
  \includegraphics[width=.95\linewidth]{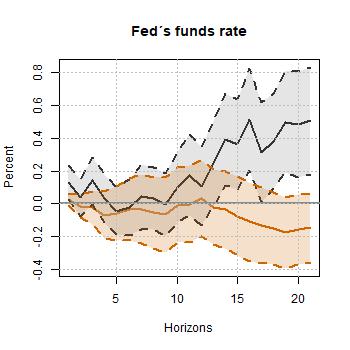}
\end{subfigure}
\begin{subfigure}{.33\textwidth}
  \centering
  \includegraphics[width=.95\linewidth]{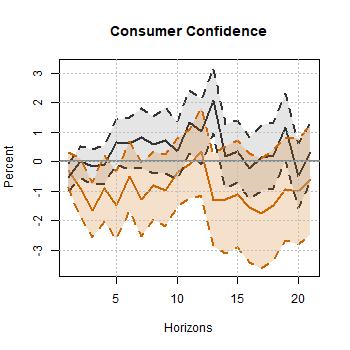}
\end{subfigure}
\begin{subfigure}{.33\textwidth}
  \centering
  \includegraphics[width=.95\linewidth]{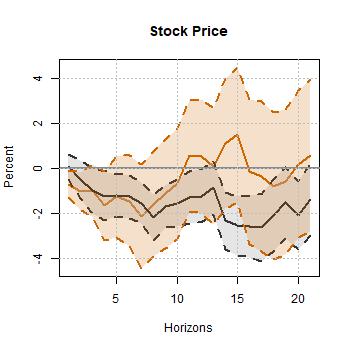}
\end{subfigure}
\hfill
\begin{subfigure}{.33\textwidth}
  \centering
  \includegraphics[width=.95\linewidth]{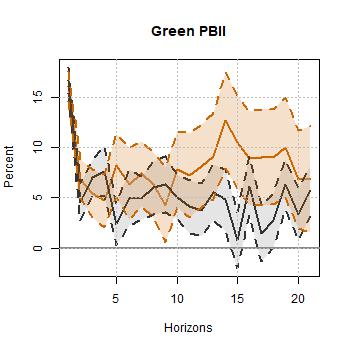}
\end{subfigure}
\begin{subfigure}{.33\textwidth}
  \centering
  \includegraphics[width=.95\linewidth]{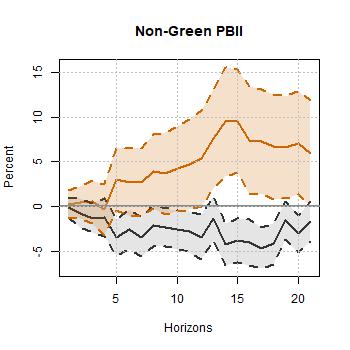}
\end{subfigure}
\caption{Dynamic responses to the idiosyncratic component of the green technology news shock. Responses before (gray) and after (orange) 1990 estimated respectively by $\beta_h(0)$ and $\beta_h(1)$ in Equation 2 using $\epsilon_{O,t}^{G}$ as the shock variable.}
\label{fig_irf_6b}
\end{figure}

Figures \ref{fig_irf_6a} and \ref{fig_irf_6b} report the estimated values and confidence bands of $\beta_h(0)$ and $\beta_h(1)$, the impulse-response conditional on being in the pre-1990 and post-1990 periods, respectively. The most salient results can be summarized as follows. First, regarding the common component, the responses of TFP and consumption are more pronounced in the second part of the sample, even though the reaction of output is little changed. The smaller impact in hours worked and consumer confidence suggest that green breakthroughs have become less unexpected for households and that a larger share of the adjustment now occurs through productivity rather than labor input.  Second, the impact of the idiosyncratic component on consumer prices is mainly observed during the pre-1990 period. A correspondingly smaller and statistically weaker policy-rate reaction is obtained. A plausible explanation is that supply-chain learning and policy predictability have reduced the marginal cost of adopting low-carbon inputs. The consumption response and the NGPBII now rise noticeably, indicating that more recent clean-tech progress diffuses faster into the broader economy.

These patterns underline that the macro impact of green innovation is state-contingent: as the clean-technology ecosystem matured, its long-run productivity dividend grew while its short-run inflation tax abated. Investigating how this time variation interacts with business-cycle conditions, policy regimes, or alternative breakpoints remains a promising avenue for future research.

\section{Conclusion} \label{conclusion}

This paper provided empirical evidence on the macroeconomic impact of the green transition through a mechanism that has been underexplored in the literature: the effects of green innovation. By tracking the market value of newly-granted green patents, we were able  to isolate two orthogonal components with differential effects on relevant macroeconomic variables. First, a technological news component, common to both green and non-green innovation, increases output and total factor productivity. The response pattern mirrors classic technology-news shocks but is identified without imposing any restriction on the TFP's responses. Second, an idiosyncratic transition component triggers a persistent rise in consumer prices and a fall in stock valuations, consistent with investors front-loading the cost of future, tighter carbon policy. These twin channels reconcile the \say{greenflation} narrative with the long-run growth upside of clean technologies.

Our findings matter for economic policy. Because transition news moves prices even when energy costs are stable, monetary authorities cannot treat the green shift as a standard supply shock. Forward-looking communication and credible carbon-policy road-maps could blunt the inflationary leg of transition news while preserving its innovation stimulus. Fiscal tools that speed up energy-efficiency upgrades likewise dampen the price pass-through we document. Ignoring these interactions risks a monetary–climate policy tug-of-war. For researchers, our patent-valuation strategy opens interesting research fronts. Cross-country replications can test whether institutional differences temper the inflation–growth trade-off we uncover. Linking the shock series to firm-level balance sheets would reveal winners and losers inside each sector. Finally, our focus on green innovation can be extended to understand the effect of other domains of technology-specific news shocks. An interesting line of research in this direction would be to explore more deeply time-varying effects in productivity.

\newpage

\bibliography{References}

\begin{thebibliography}{}

\bibitem [\protect \citeauthoryear {%
Acemoglu%
, Aghion%
, Bursztyn%
\BCBL {}\ \BBA {} Hemous%
}{%
Acemoglu%
\ \protect \BOthers {.}}{%
{\protect \APACyear {2012}}%
}]{%
Acemogluetal2012}
\APACinsertmetastar {%
Acemogluetal2012}%
\begin{APACrefauthors}%
Acemoglu, D.%
, Aghion, P.%
, Bursztyn, L.%
\BCBL {}\ \BBA {} Hemous, D.%
\end{APACrefauthors}%
\unskip\
\newblock
\APACrefYearMonthDay{2012}{}{}.
\newblock
{\BBOQ}\APACrefatitle {{The Environment and Directed Technical Change}} {{The Environment and Directed Technical Change}}.{\BBCQ}
\newblock
\APACjournalVolNumPages{American Economic Review}{102}{1}{131-66}.
\newblock
\begin{APACrefDOI} \doi{10.1257/aer.102.1.131} \end{APACrefDOI}
\PrintBackRefs{\CurrentBib}

\bibitem [\protect \citeauthoryear {%
Acemoglu%
, Akcigit%
, Hanley%
\BCBL {}\ \BBA {} Kerr%
}{%
Acemoglu%
\ \protect \BOthers {.}}{%
{\protect \APACyear {2016}}%
}]{%
Acemogluetal2016}
\APACinsertmetastar {%
Acemogluetal2016}%
\begin{APACrefauthors}%
Acemoglu, D.%
, Akcigit, U.%
, Hanley, D.%
\BCBL {}\ \BBA {} Kerr, W.%
\end{APACrefauthors}%
\unskip\
\newblock
\APACrefYearMonthDay{2016}{}{}.
\newblock
{\BBOQ}\APACrefatitle {{Transition to Clean Technology}} {{Transition to Clean Technology}}.{\BBCQ}
\newblock
\APACjournalVolNumPages{Journal of Political Economy}{124}{1}{52-104}.
\newblock
\begin{APACrefDOI} \doi{10.1086/684511} \end{APACrefDOI}
\PrintBackRefs{\CurrentBib}

\bibitem [\protect \citeauthoryear {%
Aghion%
, Dechezlepr\^{e}tre%
, H\'{e}mous%
, Martin%
\BCBL {}\ \BBA {} Van~Reenen%
}{%
Aghion%
\ \protect \BOthers {.}}{%
{\protect \APACyear {2016}}%
}]{%
Aghionetal2016}
\APACinsertmetastar {%
Aghionetal2016}%
\begin{APACrefauthors}%
Aghion, P.%
, Dechezlepr\^{e}tre, A.%
, H\'{e}mous, D.%
, Martin, R.%
\BCBL {}\ \BBA {} Van~Reenen, J.%
\end{APACrefauthors}%
\unskip\
\newblock
\APACrefYearMonthDay{2016}{}{}.
\newblock
{\BBOQ}\APACrefatitle {{Carbon Taxes, Path Dependency, and Directed Technical Change: Evidence from the Auto Industry}} {{Carbon Taxes, Path Dependency, and Directed Technical Change: Evidence from the Auto Industry}}.{\BBCQ}
\newblock
\APACjournalVolNumPages{Journal of Political Economy}{124}{1}{1-51}.
\newblock
\begin{APACrefDOI} \doi{10.1086/684581} \end{APACrefDOI}
\PrintBackRefs{\CurrentBib}

\bibitem [\protect \citeauthoryear {%
Airaudo%
, Pappa%
\BCBL {}\ \BBA {} Seoane%
}{%
Airaudo%
\ \protect \BOthers {.}}{%
{\protect \APACyear {2024}}%
}]{%
Airaudoetal}
\APACinsertmetastar {%
Airaudoetal}%
\begin{APACrefauthors}%
Airaudo, F.%
, Pappa, E.%
\BCBL {}\ \BBA {} Seoane, H.%
\end{APACrefauthors}%
\unskip\
\newblock
\APACrefYearMonthDay{2024}{}{}.
\newblock
\APACrefbtitle {{The Green Metamorphosis of a Small Open Economy}} {{The Green Metamorphosis of a Small Open Economy}}\ \APACbVolEdTR {}{Working Paper}.
\PrintBackRefs{\CurrentBib}

\bibitem [\protect \citeauthoryear {%
Ambec%
\ \BBA {} Lanoie%
}{%
Ambec%
\ \BBA {} Lanoie%
}{%
{\protect \APACyear {2008}}%
}]{%
AmbecLanoie}
\APACinsertmetastar {%
AmbecLanoie}%
\begin{APACrefauthors}%
Ambec, S.%
\BCBT {}\ \BBA {} Lanoie, P.%
\end{APACrefauthors}%
\unskip\
\newblock
\APACrefYearMonthDay{2008}{}{}.
\newblock
{\BBOQ}\APACrefatitle {{Does It Pay to Be Green? A Systematic Overview}} {{Does It Pay to Be Green? A Systematic Overview}}.{\BBCQ}
\newblock
\APACjournalVolNumPages{Academy of Management Perspectives}{22}{4}{45--62}.
\PrintBackRefs{\CurrentBib}

\bibitem [\protect \citeauthoryear {%
Antoine~Dechezlepretre%
}{%
Antoine~Dechezlepretre%
}{%
{\protect \APACyear {2017}}%
}]{%
Dechezlepretre}
\APACinsertmetastar {%
Dechezlepretre}%
\begin{APACrefauthors}%
Antoine~Dechezlepretre, M\BPBI M., Ralf~Martin.%
\end{APACrefauthors}%
\unskip\
\newblock
\APACrefYearMonthDay{2017}{}{}.
\newblock
\APACrefbtitle {{Knowledge Spillovers from clean and dirty technologies}} {{Knowledge Spillovers from clean and dirty technologies}}\ \APACbVolEdTR {}{GRI Working Papers\ \BNUM~135}.
\newblock
\APACaddressInstitution{}{Grantham Research Institute on Climate Change and the Environment}.
\newblock
\begin{APACrefURL} \url{https://ideas.repec.org/p/lsg/lsgwps/wp135.html} \end{APACrefURL}
\PrintBackRefs{\CurrentBib}

\bibitem [\protect \citeauthoryear {%
Barsky%
\ \BBA {} Sims%
}{%
Barsky%
\ \BBA {} Sims%
}{%
{\protect \APACyear {2011}}%
}]{%
BarskySims}
\APACinsertmetastar {%
BarskySims}%
\begin{APACrefauthors}%
Barsky, R.%
\BCBT {}\ \BBA {} Sims, E.%
\end{APACrefauthors}%
\unskip\
\newblock
\APACrefYearMonthDay{2011}{}{}.
\newblock
{\BBOQ}\APACrefatitle {{News Shocks and Business Cycles}} {{News Shocks and Business Cycles}}.{\BBCQ}
\newblock
\APACjournalVolNumPages{Journal of Monetary Economics}{58}{3}{273-289}.
\PrintBackRefs{\CurrentBib}

\bibitem [\protect \citeauthoryear {%
Bartocci%
, Notarpietro%
\BCBL {}\ \BBA {} Pisani%
}{%
Bartocci%
\ \protect \BOthers {.}}{%
{\protect \APACyear {2022}}%
}]{%
Bartocci}
\APACinsertmetastar {%
Bartocci}%
\begin{APACrefauthors}%
Bartocci, A.%
, Notarpietro, A.%
\BCBL {}\ \BBA {} Pisani, M.%
\end{APACrefauthors}%
\unskip\
\newblock
\APACrefYearMonthDay{2022}{}{}.
\newblock
\APACrefbtitle {{“Green” Fiscal Policy Measures and Non-standard Monetary Policy in the Euro Area}} {{“Green” Fiscal Policy Measures and Non-standard Monetary Policy in the Euro Area}}\ \APACbVolEdTR {}{Working Paper\ \BNUM\ 1377}.
\newblock
\APACaddressInstitution{}{Bank of Italy}.
\PrintBackRefs{\CurrentBib}

\bibitem [\protect \citeauthoryear {%
Beaudry%
\ \BBA {} Portier%
}{%
Beaudry%
\ \BBA {} Portier%
}{%
{\protect \APACyear {2006}}%
}]{%
BeaudryPortier}
\APACinsertmetastar {%
BeaudryPortier}%
\begin{APACrefauthors}%
Beaudry, P.%
\BCBT {}\ \BBA {} Portier, F.%
\end{APACrefauthors}%
\unskip\
\newblock
\APACrefYearMonthDay{2006}{}{}.
\newblock
{\BBOQ}\APACrefatitle {{Stock Prices, News, and Economic Fluctuations}} {{Stock Prices, News, and Economic Fluctuations}}.{\BBCQ}
\newblock
\APACjournalVolNumPages{American Economic Review}{96}{4}{1293-1307}.
\PrintBackRefs{\CurrentBib}

\bibitem [\protect \citeauthoryear {%
Bernard%
\ \BBA {} Kichian%
}{%
Bernard%
\ \BBA {} Kichian%
}{%
{\protect \APACyear {2021}}%
}]{%
BernardKichian}
\APACinsertmetastar {%
BernardKichian}%
\begin{APACrefauthors}%
Bernard, J\BHBI T.%
\BCBT {}\ \BBA {} Kichian, M.%
\end{APACrefauthors}%
\unskip\
\newblock
\APACrefYearMonthDay{2021}{}{}.
\newblock
{\BBOQ}\APACrefatitle {{The Impact of a Revenue-Neutral Carbon Tax on GDP Dynamics: The Case of British Columbia}} {{The Impact of a Revenue-Neutral Carbon Tax on GDP Dynamics: The Case of British Columbia}}.{\BBCQ}
\newblock
\APACjournalVolNumPages{The Energy Journal}{42}{3}{205-224}.
\newblock
\begin{APACrefDOI} \doi{10.5547/01956574.42.3.jber} \end{APACrefDOI}
\PrintBackRefs{\CurrentBib}

\bibitem [\protect \citeauthoryear {%
Botta%
\ \BBA {} Ko{\'z}luk%
}{%
Botta%
\ \BBA {} Ko{\'z}luk%
}{%
{\protect \APACyear {2014}}%
}]{%
botta}
\APACinsertmetastar {%
botta}%
\begin{APACrefauthors}%
Botta, E.%
\BCBT {}\ \BBA {} Ko{\'z}luk, T.%
\end{APACrefauthors}%
\unskip\
\newblock
\APACrefYearMonthDay{2014}{}{}.
\newblock
{\BBOQ}\APACrefatitle {{Measuring Environmental Policy Stringency in OECD Countries: A Composite Index Approach}} {{Measuring Environmental Policy Stringency in OECD Countries: A Composite Index Approach}}.{\BBCQ}
\newblock

\PrintBackRefs{\CurrentBib}

\bibitem [\protect \citeauthoryear {%
Cascaldi-García%
\ \BBA {} Vukotic%
}{%
Cascaldi-García%
\ \BBA {} Vukotic%
}{%
{\protect \APACyear {2022}}%
}]{%
CascaldiGarcia}
\APACinsertmetastar {%
CascaldiGarcia}%
\begin{APACrefauthors}%
Cascaldi-García, D.%
\BCBT {}\ \BBA {} Vukotic, M.%
\end{APACrefauthors}%
\unskip\
\newblock
\APACrefYearMonthDay{2022}{}{}.
\newblock
{\BBOQ}\APACrefatitle {{Patent-Based News Shocks}} {{Patent-Based News Shocks}}.{\BBCQ}
\newblock
\APACjournalVolNumPages{The Review of Economics and Statistics}{104}{1}{51-66}.
\PrintBackRefs{\CurrentBib}

\bibitem [\protect \citeauthoryear {%
Del~Negro%
, di Giovanni%
\BCBL {}\ \BBA {} Dogra%
}{%
Del~Negro%
\ \protect \BOthers {.}}{%
{\protect \APACyear {2023}}%
}]{%
DelNegro}
\APACinsertmetastar {%
DelNegro}%
\begin{APACrefauthors}%
Del~Negro, M.%
, di Giovanni, J.%
\BCBL {}\ \BBA {} Dogra, K.%
\end{APACrefauthors}%
\unskip\
\newblock
\APACrefYearMonthDay{2023}{}{}.
\newblock
\APACrefbtitle {{Is the Green Transition Inflationary?}} {{Is the Green Transition Inflationary?}}\ \APACbVolEdTR {}{Staff Reports\ \BNUM\ 1053}.
\newblock
\APACaddressInstitution{}{Federal Reserve Bank of New York}.
\PrintBackRefs{\CurrentBib}

\bibitem [\protect \citeauthoryear {%
Fernald%
}{%
Fernald%
}{%
{\protect \APACyear {2014}}%
}]{%
Fernald}
\APACinsertmetastar {%
Fernald}%
\begin{APACrefauthors}%
Fernald, J.%
\end{APACrefauthors}%
\unskip\
\newblock
\APACrefYearMonthDay{2014}{}{}.
\newblock
{\BBOQ}\APACrefatitle {{A Quarterly Utilization-Adjusted Series on Total Factor Productivity}} {{A Quarterly Utilization-Adjusted Series on Total Factor Productivity}}.{\BBCQ}
\newblock
\APACjournalVolNumPages{Working Paper Series 2012-19, Federal Reserve Bank of San Francisco}{}{}{}.
\PrintBackRefs{\CurrentBib}

\bibitem [\protect \citeauthoryear {%
Ferrari%
\ \BBA {} Nispi~Landi%
}{%
Ferrari%
\ \BBA {} Nispi~Landi%
}{%
{\protect \APACyear {2022}}%
}]{%
Ferrari2022}
\APACinsertmetastar {%
Ferrari2022}%
\begin{APACrefauthors}%
Ferrari, A.%
\BCBT {}\ \BBA {} Nispi~Landi, V.%
\end{APACrefauthors}%
\unskip\
\newblock
\APACrefYearMonthDay{2022}{}{}.
\newblock
\APACrefbtitle {{Will the Green Transition Be Inflationary? Expectations Matter}} {{Will the Green Transition Be Inflationary? Expectations Matter}}\ \APACbVolEdTR {}{Working Paper Series\ \BNUM\ 2726}.
\newblock
\APACaddressInstitution{}{European Central Bank}.
\PrintBackRefs{\CurrentBib}

\bibitem [\protect \citeauthoryear {%
Ferrari%
\ \BBA {} Nispi~Landi%
}{%
Ferrari%
\ \BBA {} Nispi~Landi%
}{%
{\protect \APACyear {2023}}%
}]{%
Ferrari2023}
\APACinsertmetastar {%
Ferrari2023}%
\begin{APACrefauthors}%
Ferrari, A.%
\BCBT {}\ \BBA {} Nispi~Landi, V.%
\end{APACrefauthors}%
\unskip\
\newblock
\APACrefYearMonthDay{2023}{}{}.
\newblock
\APACrefbtitle {{Toward a Green Economy: The Role of Central Bank’s Asset Purchases}} {{Toward a Green Economy: The Role of Central Bank’s Asset Purchases}}\ \APACbVolEdTR {}{Working Paper Series\ \BNUM\ 2779}.
\newblock
\APACaddressInstitution{}{European Central Bank}.
\PrintBackRefs{\CurrentBib}

\bibitem [\protect \citeauthoryear {%
Francis%
, Owyang%
, Roush%
\BCBL {}\ \BBA {} DiCecio%
}{%
Francis%
\ \protect \BOthers {.}}{%
{\protect \APACyear {2014}}%
}]{%
Francisetal}
\APACinsertmetastar {%
Francisetal}%
\begin{APACrefauthors}%
Francis, N.%
, Owyang, M\BPBI T.%
, Roush, J\BPBI E.%
\BCBL {}\ \BBA {} DiCecio, R.%
\end{APACrefauthors}%
\unskip\
\newblock
\APACrefYearMonthDay{2014}{}{}.
\newblock
{\BBOQ}\APACrefatitle {{A Flexible Finite-horizon Alternative to Long-run Restrictions with an Application to Technology Shocks}} {{A Flexible Finite-horizon Alternative to Long-run Restrictions with an Application to Technology Shocks}}.{\BBCQ}
\newblock
\APACjournalVolNumPages{The Review of Economics and Statistics}{96}{4}{638--647}.
\PrintBackRefs{\CurrentBib}

\bibitem [\protect \citeauthoryear {%
Fried%
}{%
Fried%
}{%
{\protect \APACyear {2018}}%
}]{%
Fried}
\APACinsertmetastar {%
Fried}%
\begin{APACrefauthors}%
Fried, S.%
\end{APACrefauthors}%
\unskip\
\newblock
\APACrefYearMonthDay{2018}{}{}.
\newblock
{\BBOQ}\APACrefatitle {{Climate Policy and Innovation: A Quantitative Macroeconomic Analysis}} {{Climate Policy and Innovation: A Quantitative Macroeconomic Analysis}}.{\BBCQ}
\newblock
\APACjournalVolNumPages{American Economic Journal: Macroeconomics}{10}{1}{90-118}.
\newblock
\begin{APACrefURL} \url{https://www.aeaweb.org/articles?id=10.1257/mac.20150289} \end{APACrefURL}
\newblock
\begin{APACrefDOI} \doi{10.1257/mac.20150289} \end{APACrefDOI}
\PrintBackRefs{\CurrentBib}

\bibitem [\protect \citeauthoryear {%
Gavriilidis%
}{%
Gavriilidis%
}{%
{\protect \APACyear {2021}}%
}]{%
gavriilidis}
\APACinsertmetastar {%
gavriilidis}%
\begin{APACrefauthors}%
Gavriilidis, K.%
\end{APACrefauthors}%
\unskip\
\newblock
\APACrefYearMonthDay{2021}{}{}.
\newblock
{\BBOQ}\APACrefatitle {{Measuring Climate Policy Uncertainty}} {{Measuring Climate Policy Uncertainty}}.{\BBCQ}
\newblock
\APACjournalVolNumPages{Available at SSRN 3847388}{}{}{}.
\PrintBackRefs{\CurrentBib}

\bibitem [\protect \citeauthoryear {%
Goulder%
\ \BBA {} Hafstead%
}{%
Goulder%
\ \BBA {} Hafstead%
}{%
{\protect \APACyear {2018}}%
}]{%
Goulder}
\APACinsertmetastar {%
Goulder}%
\begin{APACrefauthors}%
Goulder, L\BPBI H.%
\BCBT {}\ \BBA {} Hafstead, M\BPBI A\BPBI C.%
\end{APACrefauthors}%
\unskip\
\newblock
\APACrefYear{2018}.
\newblock
\APACrefbtitle {{Confronting the Climate Challenge: U.S. Policy Options}} {{Confronting the Climate Challenge: U.S. Policy Options}}.
\newblock
\APACaddressPublisher{}{Columbia University Press}.
\PrintBackRefs{\CurrentBib}

\bibitem [\protect \citeauthoryear {%
Hasna%
, Jaumotte%
, Kim%
, Pienknagura%
\BCBL {}\ \BBA {} Schwerhoff%
}{%
Hasna%
\ \protect \BOthers {.}}{%
{\protect \APACyear {2023}}%
}]{%
Hasnaetal}
\APACinsertmetastar {%
Hasnaetal}%
\begin{APACrefauthors}%
Hasna, Z.%
, Jaumotte, F.%
, Kim, J.%
, Pienknagura, S.%
\BCBL {}\ \BBA {} Schwerhoff, G.%
\end{APACrefauthors}%
\unskip\
\newblock
\APACrefYearMonthDay{2023}{}{}.
\newblock
\APACrefbtitle {{Green Innovation and Diffusion: Policies to Accelerate Them and Expected Impact on Macroeconomic and Firm-Level Performance}} {{Green Innovation and Diffusion: Policies to Accelerate Them and Expected Impact on Macroeconomic and Firm-Level Performance}}\ \APACbVolEdTR{}{\BTR{}}.
\newblock
\APACaddressInstitution{}{Staff Discussion Note SDN/2023/008. International Monetary Fund}.
\PrintBackRefs{\CurrentBib}

\bibitem [\protect \citeauthoryear {%
Haščič%
\ \BBA {} Migotto%
}{%
Haščič%
\ \BBA {} Migotto%
}{%
{\protect \APACyear {2015}}%
}]{%
HascicMigotto}
\APACinsertmetastar {%
HascicMigotto}%
\begin{APACrefauthors}%
Haščič, I.%
\BCBT {}\ \BBA {} Migotto, M.%
\end{APACrefauthors}%
\unskip\
\newblock
\APACrefYearMonthDay{2015}{}{}.
\newblock
{\BBOQ}\APACrefatitle {{Measuring Environmental Innovation Using Patent Data}} {{Measuring Environmental Innovation Using Patent Data}}.{\BBCQ}
\newblock
\APACjournalVolNumPages{OECD Environment Working Papers No. 89}{}{}{}.
\PrintBackRefs{\CurrentBib}

\bibitem [\protect \citeauthoryear {%
Jord{\`a}%
}{%
Jord{\`a}%
}{%
{\protect \APACyear {2023}}%
}]{%
jorda2023}
\APACinsertmetastar {%
jorda2023}%
\begin{APACrefauthors}%
Jord{\`a}, {\`O}.%
\end{APACrefauthors}%
\unskip\
\newblock
\APACrefYearMonthDay{2023}{}{}.
\newblock
{\BBOQ}\APACrefatitle {Local projections for applied economics} {Local projections for applied economics}.{\BBCQ}
\newblock
\APACjournalVolNumPages{Annual Review of Economics}{15}{1}{607--631}.
\PrintBackRefs{\CurrentBib}

\bibitem [\protect \citeauthoryear {%
Jordà%
}{%
Jordà%
}{%
{\protect \APACyear {2005}}%
}]{%
Jorda}
\APACinsertmetastar {%
Jorda}%
\begin{APACrefauthors}%
Jordà, .%
\end{APACrefauthors}%
\unskip\
\newblock
\APACrefYearMonthDay{2005}{}{}.
\newblock
{\BBOQ}\APACrefatitle {{Estimation and Inference of Impulse Responses by Local Projections}} {{Estimation and Inference of Impulse Responses by Local Projections}}.{\BBCQ}
\newblock
\APACjournalVolNumPages{American Economic Review}{95}{1}{161-182}.
\newblock
\begin{APACrefURL} \url{https://www.aeaweb.org/articles?id=10.1257/0002828053828518} \end{APACrefURL}
\newblock
\begin{APACrefDOI} \doi{10.1257/0002828053828518} \end{APACrefDOI}
\PrintBackRefs{\CurrentBib}

\bibitem [\protect \citeauthoryear {%
Kogan%
, Papanikolaou%
, Seru%
\BCBL {}\ \BBA {} Stoffman%
}{%
Kogan%
\ \protect \BOthers {.}}{%
{\protect \APACyear {2017}}%
}]{%
KPSS}
\APACinsertmetastar {%
KPSS}%
\begin{APACrefauthors}%
Kogan, L.%
, Papanikolaou, D.%
, Seru, A.%
\BCBL {}\ \BBA {} Stoffman, N.%
\end{APACrefauthors}%
\unskip\
\newblock
\APACrefYearMonthDay{2017}{}{}.
\newblock
{\BBOQ}\APACrefatitle {{Technological Innovation, Resource Allocation, and Growth}} {{Technological Innovation, Resource Allocation, and Growth}}.{\BBCQ}
\newblock
\APACjournalVolNumPages{The Quarterly Journal of Economics}{132}{2}{665–712}.
\PrintBackRefs{\CurrentBib}

\bibitem [\protect \citeauthoryear {%
Kurmann%
\ \BBA {} Sims%
}{%
Kurmann%
\ \BBA {} Sims%
}{%
{\protect \APACyear {2021}}%
}]{%
KurmannSims}
\APACinsertmetastar {%
KurmannSims}%
\begin{APACrefauthors}%
Kurmann, A.%
\BCBT {}\ \BBA {} Sims, E.%
\end{APACrefauthors}%
\unskip\
\newblock
\APACrefYearMonthDay{2021}{}{}.
\newblock
{\BBOQ}\APACrefatitle {{Revisions in Utilization-Adjusted TFP and Robust Identification of News Shocks}} {{Revisions in Utilization-Adjusted TFP and Robust Identification of News Shocks}}.{\BBCQ}
\newblock
\APACjournalVolNumPages{The Review of Economics and Statistics}{103}{2}{216-235}.
\newblock
\begin{APACrefDOI} \doi{10.1162/rest_a_00896} \end{APACrefDOI}
\PrintBackRefs{\CurrentBib}

\bibitem [\protect \citeauthoryear {%
Känzig%
}{%
Känzig%
}{%
{\protect \APACyear {2023}}%
}]{%
Kanzing}
\APACinsertmetastar {%
Kanzing}%
\begin{APACrefauthors}%
Känzig, D\BPBI R.%
\end{APACrefauthors}%
\unskip\
\newblock
\APACrefYearMonthDay{2023}{}{}.
\newblock
\APACrefbtitle {The Unequal Economic Consequences of Carbon Pricing} {The unequal economic consequences of carbon pricing}\ \APACbVolEdTR {}{Working Paper\ \BNUM\ 31221}.
\newblock
\APACaddressInstitution{}{National Bureau of Economic Research}.
\newblock
\begin{APACrefDOI} \doi{10.3386/w31221} \end{APACrefDOI}
\PrintBackRefs{\CurrentBib}

\bibitem [\protect \citeauthoryear {%
G.~Metcalf%
}{%
G.~Metcalf%
}{%
{\protect \APACyear {2019}}%
}]{%
Metcalf}
\APACinsertmetastar {%
Metcalf}%
\begin{APACrefauthors}%
Metcalf, G.%
\end{APACrefauthors}%
\unskip\
\newblock
\APACrefYearMonthDay{2019}{}{}.
\newblock
{\BBOQ}\APACrefatitle {{On the Economics of a Carbon Tax for the United States}} {{On the Economics of a Carbon Tax for the United States}}.{\BBCQ}
\newblock
\APACjournalVolNumPages{Brookings Papers on Economic Activity}{50}{1}{405-458}.
\PrintBackRefs{\CurrentBib}

\bibitem [\protect \citeauthoryear {%
G\BPBI E.~Metcalf%
\ \BBA {} Stock%
}{%
G\BPBI E.~Metcalf%
\ \BBA {} Stock%
}{%
{\protect \APACyear {2023}}%
}]{%
MetcalfStock}
\APACinsertmetastar {%
MetcalfStock}%
\begin{APACrefauthors}%
Metcalf, G\BPBI E.%
\BCBT {}\ \BBA {} Stock, J\BPBI H.%
\end{APACrefauthors}%
\unskip\
\newblock
\APACrefYearMonthDay{2023}{}{}.
\newblock
{\BBOQ}\APACrefatitle {{The Macroeconomic Impact of Europe's Carbon Taxes}} {{The Macroeconomic Impact of Europe's Carbon Taxes}}.{\BBCQ}
\newblock
\APACjournalVolNumPages{American Economic Journal: Macroeconomics}{15}{3}{265-86}.
\newblock
\begin{APACrefDOI} \doi{10.1257/mac.20210052} \end{APACrefDOI}
\PrintBackRefs{\CurrentBib}

\bibitem [\protect \citeauthoryear {%
Miranda-Agrippino%
, Hacioglu-Hoke%
\BCBL {}\ \BBA {} Bluwstein%
}{%
Miranda-Agrippino%
\ \protect \BOthers {.}}{%
{\protect \APACyear {2022}}%
}]{%
MirandaAgrippino}
\APACinsertmetastar {%
MirandaAgrippino}%
\begin{APACrefauthors}%
Miranda-Agrippino, S.%
, Hacioglu-Hoke, S.%
\BCBL {}\ \BBA {} Bluwstein, K.%
\end{APACrefauthors}%
\unskip\
\newblock
\APACrefYearMonthDay{2022}{}{}.
\newblock
{\BBOQ}\APACrefatitle {{Patents, News, and Business Cycles}} {{Patents, News, and Business Cycles}}.{\BBCQ}
\newblock
\APACjournalVolNumPages{Working Paper}{}{}{}.
\PrintBackRefs{\CurrentBib}

\bibitem [\protect \citeauthoryear {%
Sims%
, Stock%
\BCBL {}\ \BBA {} Watson%
}{%
Sims%
\ \protect \BOthers {.}}{%
{\protect \APACyear {1990}}%
}]{%
sims1990}
\APACinsertmetastar {%
sims1990}%
\begin{APACrefauthors}%
Sims, C\BPBI A.%
, Stock, J\BPBI H.%
\BCBL {}\ \BBA {} Watson, M\BPBI W.%
\end{APACrefauthors}%
\unskip\
\newblock
\APACrefYearMonthDay{1990}{}{}.
\newblock
{\BBOQ}\APACrefatitle {{Inference in Linear Time Series Models with Some Unit Roots}} {{Inference in Linear Time Series Models with Some Unit Roots}}.{\BBCQ}
\newblock
\APACjournalVolNumPages{Econometrica: Journal of the Econometric Society}{}{}{113--144}.
\PrintBackRefs{\CurrentBib}

\end{thebibliography}


@article{MirandaAgrippino,
  title="{Patents, News, and Business Cycles}",
  author={Silvia Miranda-Agrippino and Sinem Hacioglu-Hoke and Kristina Bluwstein},
  journal={Working Paper},
  year={2022},
}

@article{CascaldiGarcia,
  title="{Patent-Based News Shocks}",
  author={Danilo Cascaldi-García and Marija Vukotic},
  journal={The Review of Economics and Statistics},
  number = {1},
  volume={104},
  pages={51-66},
  year={2022},
}
@article{CascaldiGarciaZubairy,
  title="{Innovation During Challenging Times}",
  author={Danilo Cascaldi-García and Marija Vukotic and Sarah Zubairy},
  journal={Unpublished Manuscript},
  year={2023},
}
@article{HascicMigotto,
  title="{Measuring Environmental Innovation Using Patent Data}",
  author={Ivan Haščič and Mauro Migotto },
  journal={OECD Environment Working Papers No. 89},
  year={2015},
}
@article{jorda2023,
  title={Local projections for applied economics},
  author={Jord{\`a}, {\`O}scar},
  journal={Annual Review of Economics},
  volume={15},
  number={1},
  pages={607--631},
  year={2023},
  publisher={Annual Reviews}
}
@article{Fernald,
  title="{A Quarterly Utilization-Adjusted Series on Total Factor Productivity}",
  author={John Fernald},
  journal={Working Paper Series 2012-19, Federal Reserve Bank of San Francisco},
  year={2014},
}

@article{BeaudryPortier,
  title="{Stock Prices, News, and Economic Fluctuations}",
  author={Paul Beaudry and Franck Portier},
  journal={American Economic Review},
  number = {4},
  volume={96},
  pages={1293-1307},
  year={2006},
}

@article{BarskySims,
  title="{News Shocks and Business Cycles}",
  author={Robert Barsky and Eric Sims},
  journal={Journal of Monetary Economics},
  number = {3},
  volume={58},
  pages={273-289},
  year={2011},
}

@article{KPSS,
  title="{Technological Innovation, Resource Allocation, and Growth}",
  author={Leonid Kogan and Dimitris Papanikolaou and Amit Seru and Noah Stoffman},
  journal={The Quarterly Journal of Economics},
  number = {2},
  volume={132},
  pages={665–712},
  year={2017},
}

@article{sims1990,
  title="{Inference in Linear Time Series Models with Some Unit Roots}",
  author={Sims, Christopher A and Stock, James H and Watson, Mark W},
  journal={Econometrica: Journal of the Econometric Society},
  pages={113--144},
  year={1990},
  publisher={JSTOR}
}

@article{Francisetal,
 author = {Neville Francis and Michael T. Owyang and Jennifer E. Roush and Riccardo DiCecio},
 journal = {The Review of Economics and Statistics},
 number = {4},
 pages = {638--647},
 publisher = {The MIT Press},
 title = "{A Flexible Finite-horizon Alternative to Long-run Restrictions with an Application to Technology Shocks}",
volume = {96},
 year = {2014}
}

@article{KurmannSims,
    author = {Kurmann, André and Sims, Eric},
    title = "{Revisions in Utilization-Adjusted TFP and Robust Identification of News Shocks}",
    journal = {The Review of Economics and Statistics},
    volume = {103},
    number = {2},
    pages = {216-235},
    year = {2021},
    doi = {10.1162/rest_a_00896},
}
@techreport{Hasnaetal,
    author = {Zeina Hasna and Florence Jaumotte and Jaden Kim and Samuel Pienknagura and Gregor Schwerhoff},
    title = "{Green Innovation and Diffusion: Policies to Accelerate Them and Expected Impact on Macroeconomic and Firm-Level Performance}",
    institution = {Staff Discussion Note SDN/2023/008. International Monetary Fund} ,
    year = {2023},
}

@article{Acemogluetal2016,
author = {Acemoglu, Daron and Akcigit, Ufuk and Hanley, Douglas and Kerr, William},
title = "{Transition to Clean Technology}",
journal = {Journal of Political Economy},
volume = {124},
number = {1},
pages = {52-104},
year = {2016},
doi = {10.1086/684511},
}

@article{Acemogluetal2012,
Author = {Acemoglu, Daron and Aghion, Philippe and Bursztyn, Leonardo and Hemous, David},
Title = "{The Environment and Directed Technical Change}",
Journal = {American Economic Review},
Volume = {102},
Number = {1},
Year = {2012},
Pages = {131-66},
DOI = {10.1257/aer.102.1.131},
}

@book{Goulder,
 ISBN = {9780231179027},
 author = {Lawrence H. Goulder and Marc A. C. Hafstead},
 publisher = {Columbia University Press},
 title = "{Confronting the Climate Challenge: U.S. Policy Options}",
 urldate = {2024-03-08},
 year = {2018}
}

@techreport{Ferrari2023,
author = {Alessandro Ferrari and Nispi Landi, Valerio},
type = {Working Paper Series},
institution = {European Central Bank},
series = {Working Paper Series},
number = {2779},
title = "{Toward a Green Economy: The Role of Central Bank’s Asset Purchases}",
year = {2023}
}

@techreport{Bartocci,
author = {Bartocci, Anna and Notarpietro, Alessandro and Pisani, Massimiliano},
institution = {Bank of Italy},
type = {Working Paper},
series = {Temi di Discussione},
number = {1377},
title = "{“Green” Fiscal Policy Measures and Non-standard Monetary Policy in the Euro Area}",
year = {2022}
}

@techreport{Ferrari2022,
author = {Alessandro Ferrari and Nispi Landi, Valerio},
institution = {European Central Bank},
type = {Working Paper Series},
series = {Working Paper Series},
number = {2726},
title = "{Will the Green Transition Be Inflationary? Expectations Matter}",
year = {2022}
}

@techreport{Airaudoetal,
author = {Florencia Airaudo and Evi Pappa and Hernán Seoane},
type = {Working Paper},
title = "{The Green Metamorphosis of a Small Open Economy}",
year = {2024}
}

@techreport{DelNegro,
author = {Del Negro, Marco and Julian di Giovanni and Keshav Dogra},
institution = {Federal Reserve Bank of New York},
type = {Staff Reports},
series = {Staff Reports},
number = { 1053},
title = "{Is the Green Transition Inflationary?}",
year = {2023}
}

@article{Metcalf,
Author = {Metcalf, Gilbert },
Title = "{On the Economics of a Carbon Tax for the United States}",
Journal = {Brookings Papers on Economic Activity},
Volume = {50},
Number = {1},
Year = {2019},
Pages = {405-458},
}

@article{BernardKichian,
author = {Jean-Thomas Bernard and Maral Kichian},
title = "{The Impact of a Revenue-Neutral Carbon Tax on GDP Dynamics: The Case of British Columbia}",
journal = {The Energy Journal},
volume = {42},
number = {3},
pages = {205-224},
year = {2021},
doi = {10.5547/01956574.42.3.jber},
}

@article{MetcalfStock,
Author = {Metcalf, Gilbert E. and Stock, James H.},
Title = "{The Macroeconomic Impact of Europe's Carbon Taxes}",
Journal = {American Economic Journal: Macroeconomics},
Volume = {15},
Number = {3},
Year = {2023},
Pages = {265-86},
DOI = {10.1257/mac.20210052},
}


@techreport{Kanzing,
 title = "The Unequal Economic Consequences of Carbon Pricing",
 author = "Känzig, Diego R",
 institution = "National Bureau of Economic Research",
 type = "Working Paper",
 series = "Working Paper Series",
 number = "31221",
 year = "2023",
 doi = {10.3386/w31221},
}

@article{Aghionetal2016,
author = {Aghion, Philippe and Dechezlepr\^{e}tre, Antoine and H\'{e}mous, David and Martin, Ralf and Van Reenen, John},
title = "{Carbon Taxes, Path Dependency, and Directed Technical Change: Evidence from the Auto Industry}",
journal = {Journal of Political Economy},
volume = {124},
number = {1},
pages = {1-51},
year = {2016},
doi = {10.1086/684581},
}

@article{AmbecLanoie,
 author = {Stefan Ambec and Paul Lanoie},
 journal = {Academy of Management Perspectives},
 number = {4},
 pages = {45--62},
 publisher = {Academy of Management},
 title = "{Does It Pay to Be Green? A Systematic Overview}",
 volume = {22},
 year = {2008}
}

@article{Fried,
Author = {Fried, Stephie},
Title = "{Climate Policy and Innovation: A Quantitative Macroeconomic Analysis}",
Journal = {American Economic Journal: Macroeconomics},
Volume = {10},
Number = {1},
Year = {2018},
Pages = {90-118},
DOI = {10.1257/mac.20150289},
URL = {https://www.aeaweb.org/articles?id=10.1257/mac.20150289}
}

@TechReport{Dechezlepretre,
  author={Antoine Dechezlepretre, Ralf Martin, Myra Mohnen},
  title="{Knowledge Spillovers from clean and dirty technologies}",
  year={2017},
  institution={Grantham Research Institute on Climate Change and the Environment},
  type={GRI Working Papers},
  url={https://ideas.repec.org/p/lsg/lsgwps/wp135.html},
  number={135},
}

@article{Jorda,
Author = {Jordà, Òscar},
Title = "{Estimation and Inference of Impulse Responses by Local Projections}",
Journal = {American Economic Review},
Volume = {95},
Number = {1},
Year = {2005},
Pages = {161-182},
DOI = {10.1257/0002828053828518},
URL = {https://www.aeaweb.org/articles?id=10.1257/0002828053828518}}

@article{Ottonello,
author = {Ottonello, Pablo and Winberry, Thomas},
title = "{Financial Heterogeneity and the Investment Channel of Monetary Policy}",
journal = {Econometrica},
volume = {88},
number = {6},
pages = {2473-2502},
year = {2020},
}

@article{Adammer,
author = {Adämmer, Philipp},
title = "{lpirfs: An R Package to Estimate Impulse
Response Functions by Local Projections}",
journal = {The R Journal},
volume = {11},
number = {2},
year = {2019},
}

@article{NeweyWest,
author = {Newey, Whitney and West, Kenneth},
title = "{A Simple, Positive Semi-Definite, Heteroskedasticity and Autocorrelation Consistent Covariance Matrix}",
journal = {Econometrica},
volume = {55},
number = {3},
pages = {703-708},
year = {1987},
}

@article{botta,
  title="{Measuring Environmental Policy Stringency in OECD Countries: A Composite Index Approach}",
  author={Botta, Enrico and Ko{\'z}luk, Tomasz},
  year={2014},
  publisher={OECD},
}

@article{gavriilidis,
  title="{Measuring Climate Policy Uncertainty}",
  author={Gavriilidis, Konstantinos},
  journal={Available at SSRN 3847388},
  year={2021},
}

\newpage

\appendix
\setcounter{table}{0}
\renewcommand{\thetable}{A\arabic{table}}
\setcounter{figure}{0}
\renewcommand{\thefigure}{A\arabic{figure}}

\section{Appendix}

\subsection{Data Sources}

This Appendix gives more details on the sources of the data used in the empirical analysis.

\begin{table}[H]
\centering
\caption{Data description and sources}
\resizebox{\textwidth}{!}{
\begin{tabular}{lcc}
\hline \hline
\textbf{Variable} & \textbf{Description} & \textbf{Source} \\ \hline
\textit{Aggregate Analysis:} \\
GPBII and NGPBII & Green and Non-Green Patent Based Innovation index & Patents view + \cite{KPSS} \\
TFP & Total Factor Productivity & \cite{Fernald} \\
Output & U.S. Real Gross Domestic Product & FRED \\
Consumption & U.S. Real Consumption & FRED \\
Investment & U.S. Real Investment & FRED \\
Hours & Non-farm Business Sector: Hours Worked for All Workers & FRED \\
Consumer Price Index & Consumer Price Index for All Urban Consumers: All Items & FRED \\
Federal Funds Rate & Effective federal funds rate & FRED \\
Consumer Confidence & Consumer Confidence Index & Michigan Survey of Consumers \\
Stock Price Index & Standard and Poor’s 500 Composite Stock Price Index & Robert Shiller’s website\\
CPI Core & U.S. CPI for all urban consumers: all items less food
and energy & FRED \\
CPI Energy & U.S. CPI for all urban consumers: energy & FRED \\
CPI Durables & U.S. CPI for all urban consumers: durables & FRED \\
CPI Non-durables & U.S. CPI for all urban consumers: non-durables & FRED \\
CPI Services & U.S. CPI for all urban consumers: services & FRED \\
Oil and Gas stock price & Average oil and gas sectoral stock price index & Datastream \\
Electricity Stock Price & Average electricity sectoral stock price index & Datastream \\
Automobiles Stock Price & Average automobiles sectoral stock price index & Datastream \\
Mining Stock Price & Average mining sectoral stock price index & Datastream \\
Retail Stock Price & Average retail sectoral stock price index & Datastream \\
Travel and Leisure Stock Price & Average travel and leisure sectoral stock price index & Datastream \\
\hline \hline\end{tabular}}
\end{table}

\subsection{Identification Using a Choleski Rotation}

\begin{figure}[H]
\begin{subfigure}{.33\textwidth}
  \centering
  \includegraphics[width=.95\linewidth]{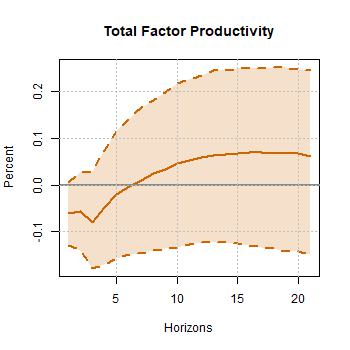}
\end{subfigure}
\begin{subfigure}{.33\textwidth}
  \centering
  \includegraphics[width=.95\linewidth]{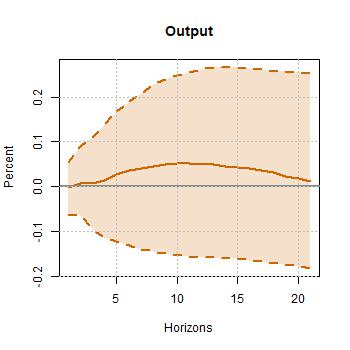}
\end{subfigure}
\begin{subfigure}{.33\textwidth}
  \centering
  \includegraphics[width=.95\linewidth]{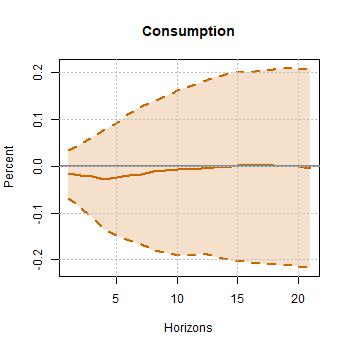}
\end{subfigure}%
\hfill
\begin{subfigure}{.33\textwidth}
  \centering
  \includegraphics[width=.95\linewidth]{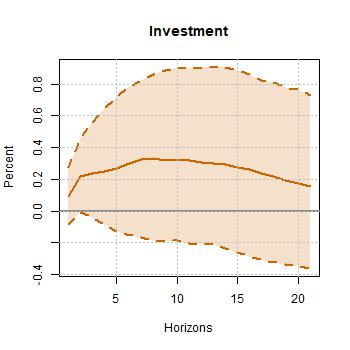}
\end{subfigure}
\begin{subfigure}{.33\textwidth}
  \centering
  \includegraphics[width=.95\linewidth]{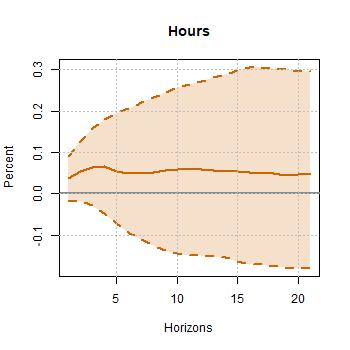}
\end{subfigure}
\begin{subfigure}{.33\textwidth}
  \centering
  \includegraphics[width=.95\linewidth]{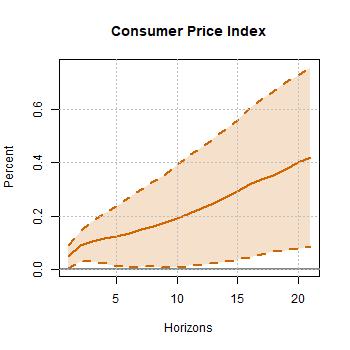}
\end{subfigure}%
\hfill
\begin{subfigure}{.33\textwidth}
  \centering
  \includegraphics[width=.95\linewidth]{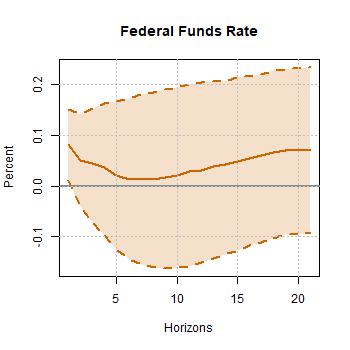}
\end{subfigure}
\begin{subfigure}{.33\textwidth}
  \centering
  \includegraphics[width=.95\linewidth]{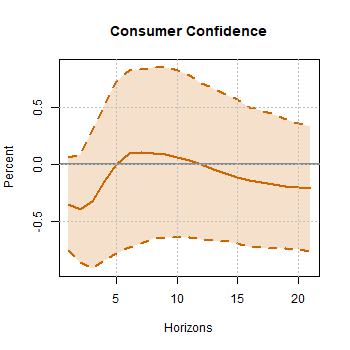}
\end{subfigure}
\begin{subfigure}{.33\textwidth}
  \centering
  \includegraphics[width=.95\linewidth]{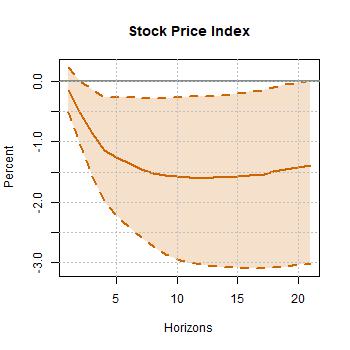}
\end{subfigure}
\hfill
\begin{subfigure}{.33\textwidth}
  \centering
  \includegraphics[width=.95\linewidth]{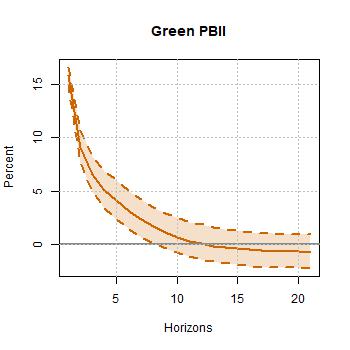}
\end{subfigure}
\begin{subfigure}{.33\textwidth}
  \centering
  \includegraphics[width=.95\linewidth]{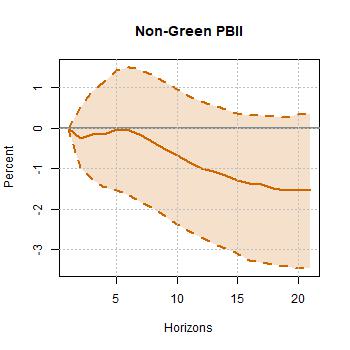}
\end{subfigure}
\caption{Dynamic responses to green technology news shocks identified using Choleski}
\label{fig_irf_choleski}
\end{figure}

The QR approach to decomposing the reduced-form residuals $e_{t}^{G}$ is equivalent to a Choleski rotation in a system where NGPBII is ordered first and GPBII is ordered second within the information set. Identification is achieved by assuming that green technology news shocks, which induce movement in the GPBII, do not affect the NGPBII on impact. This assumption suggests a higher degree of exogeneity for the non-green innovation, which is plausible if we conceive green technology innovation as nested within non-green innovation. The impulse responses to a structural green technology news shock, identified under this ordering assumption, are identical to those in Figure \ref{fig_irf_2} computed using local projections. For comparison, Figure \ref{fig_irf_choleski} presents the impulse responses from the estimated VAR model that produce similar interpretations. The decision to maintain the QR results as baseline is motivated by the possibility to obtain responses to the common projected component.

\subsection{Including Energy Efficiency in the BVAR}

\begin{figure}[H]
\begin{subfigure}{.33\textwidth}
  \centering
  \includegraphics[width=.95\linewidth]{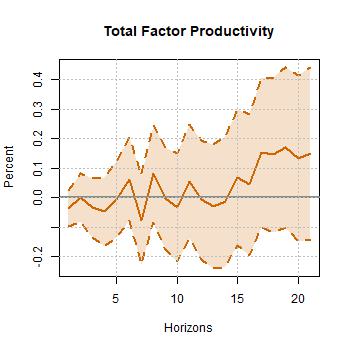}
\end{subfigure}
\begin{subfigure}{.33\textwidth}
  \centering
  \includegraphics[width=.95\linewidth]{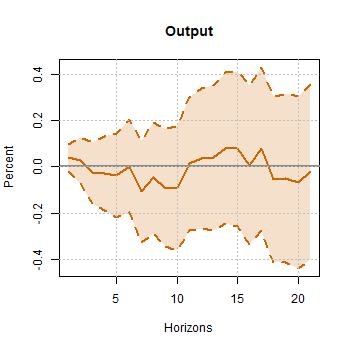}
\end{subfigure}
\begin{subfigure}{.33\textwidth}
  \centering
  \includegraphics[width=.95\linewidth]{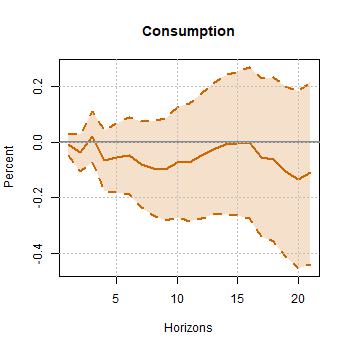}
\end{subfigure}%
\hfill
\begin{subfigure}{.33\textwidth}
  \centering
  \includegraphics[width=.95\linewidth]{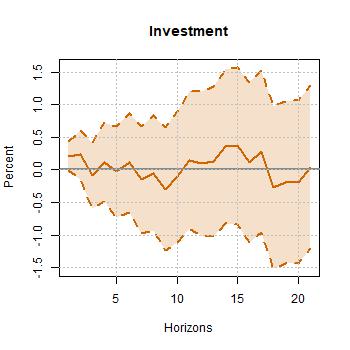}
\end{subfigure}
\begin{subfigure}{.33\textwidth}
  \centering
  \includegraphics[width=.95\linewidth]{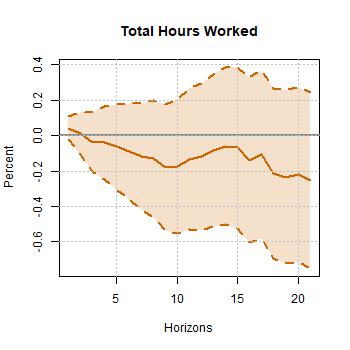}
\end{subfigure}
\begin{subfigure}{.33\textwidth}
  \centering
  \includegraphics[width=.95\linewidth]{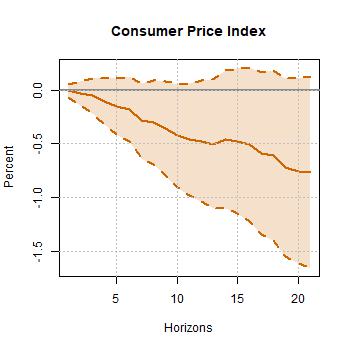}
\end{subfigure}%
\hfill
\begin{subfigure}{.33\textwidth}
  \centering
  \includegraphics[width=.95\linewidth]{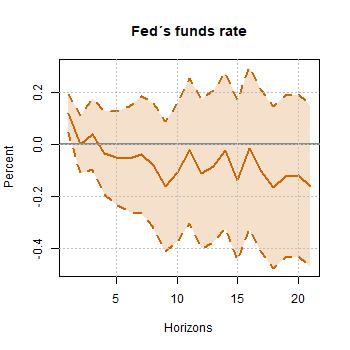}
\end{subfigure}
\begin{subfigure}{.33\textwidth}
  \centering
  \includegraphics[width=.95\linewidth]{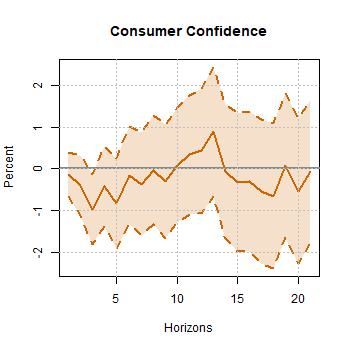}
\end{subfigure}
\begin{subfigure}{.33\textwidth}
  \centering
  \includegraphics[width=.95\linewidth]{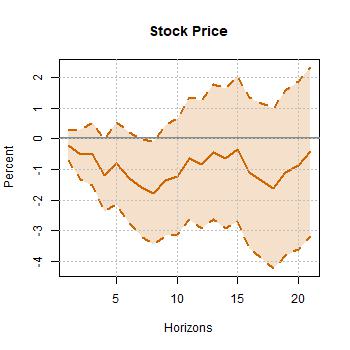}
\end{subfigure}
\hfill
\begin{subfigure}{.33\textwidth}
  \centering
  \includegraphics[width=.95\linewidth]{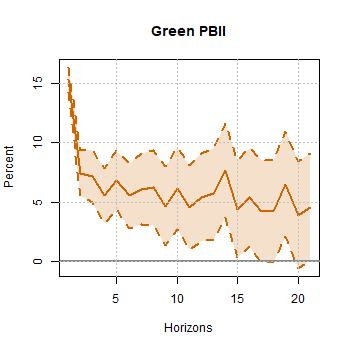}
\end{subfigure}
\begin{subfigure}{.33\textwidth}
  \centering
  \includegraphics[width=.95\linewidth]{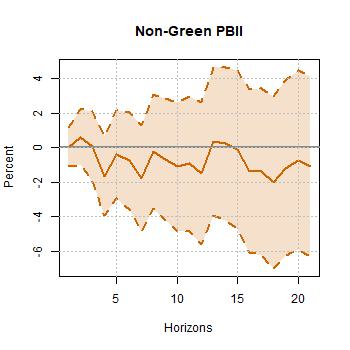}
\end{subfigure}
\caption{Dynamic responses to the idiosyncratic transition content of green technology news including the energy efficiency in the BVAR system.}
\label{fig_irf_A2}
\end{figure}

\end{document}